\documentclass[aps,pra,reprint,superscriptaddress,floatfix]{revtex4-1}

\usepackage{graphicx}
\usepackage{amsmath}
\usepackage{amssymb}

\providecommand{\ket}[1]{\ensuremath{\left|{#1}\right.\rangle}}
\providecommand{\bra}[1]{\ensuremath{\langle\left.{#1}\right|}}

\providecommand{\innerprod}[2]{\ensuremath{ \left\langle #1 | #2 \right\rangle}}

\begin{document}

\title{Few-particle dynamics of fractional quantum Hall lattice models}

\author{Dillip K. Nandy}
\affiliation{Department of Physics and Astronomy, Aarhus University, DK-8000 Aarhus C, Denmark}
\affiliation{Institute of Physics, Polish Academy of Sciences, Aleja Lotnikow 32/46, PL-02668 Warsaw, Poland}

\author{Masudul Haque}
\affiliation{Department of Theoretical Physics, Maynooth University, Co.\ Kildare, Ireland}
\affiliation{Max-Planck-Institut f\"{u}r Physik komplexer Systeme, D-01187 Dresden, Germany}

\author{Anne E. B. Nielsen}
\altaffiliation{On leave from Department of Physics and Astronomy, Aarhus University, DK-8000 Aarhus C, Denmark}
\affiliation{Max-Planck-Institut f\"{u}r Physik komplexer Systeme, D-01187 Dresden, Germany}

\begin{abstract}
Considering lattice Hamiltonians designed using conformal field theory to have fractional quantum Hall states as ground states, we study the dynamics of one or two particles on such lattices. Examining the eigenspectrum and dynamics of the single-particle sector, we demonstrate that these Hamiltonians cannot be regarded as describing interacting particles placed on a Chern band, so that the physics is fundamentally different from fractional Chern insulators. The single-particle spectrum is shown to consist of eigenstates localized in shells which have larger radius for larger eigenenergies. This leads to chiral dynamics along reasonably well-defined orbits, both in the single-particle and in the two-particle sectors.
\end{abstract}

\maketitle

\section{Introduction}

Fractional quantum Hall (FQH) states such as the Laughlin state~\cite{Laughlin_PRL1983} are widely considered to be fascinating phases of matter, particularly because they are the primary experimental realization of topological order~\cite{Wen_PRB1989,Wen_IJMP1990,WenNiu_PRB1990}.
Ever since their realization with 2D electron gases, mechanisms have been proposed for realizing FQH states in various other platforms.  Theoretically proposed lattice versions of FQH states include chiral spin liquids~\cite{KalmeyerLaughlin_PRL1987, Thomale_Greiter_chiralspinliquid_PRL2007,
Yao_Kivelson_chiralspinliquid_PRL2007}, the ground state of interacting bosons in the presence of an effective magnetic field \cite{SorensenDemlerLukin_PRL2005,HafeziSorensenDemlerLukin_PRA2007}, and fractional Chern insulators~\cite{NeupertChamonMudry_FCI_PRL2011,RegnaultBernevig_PRX2011}.
In this work we are interested in another class of lattice Hamiltonians which have FQH states as ground states --- namely, those derived in the framework of conformal field theory (CFT)
\cite{Tu_Nielsen_Cirac_Sierra_2014}.
These Hamiltonians are defined on open-boundary lattices of essentially any shape \cite{Glasser_Cirac_Sierra_Nielsen_2016}.  They involve few-body, but long-distance, terms, and they have analytical ground states. Apart from that, very little is currently known about them.

In this work, we consider the one-particle and two-particle dynamics on such lattice Hamiltonians. Specifically, we take lattice Hamiltonians which would generate a fermionic Laughlin state if the lattice filling were $1/3$, or a bosonic Laughlin state if the lattice filling were $1/2$.  Instead of these fillings, we consider either a single particle or two particles and consider the dynamics and eigenspectrum of such systems.  For the single-particle case, there is no difference between fermions and bosons.  For the two-particle case, we consider both the fermionic case and the (hard-core) bosonic case.

The motivation for this study is twofold.  Firstly, the specific motivation is to obtain additional insights into this very special class of models.  Indeed, our study of the structure and dynamics of the single-particle sector teaches us that these Hamiltonians cannot be regarded as a combination of Chern bands plus interactions.  This contrasts sharply to the mechanism of obtaining FQH-like states in fractional Chern insulators.  Our results on the two-particle systems reveal the extent to which interactions affect the spectrum and dynamics.

Secondly, a more general motivation is that the dynamics of quantum particles in topological backgrounds has been of increasing interest in the last few years.  This activity lies at the intersection of two prominent subfields of 21st-century condensed matter physics: topological matter and non-equilibrium dynamics.  Recently, some real-time dynamical aspects of interacting particles in Landau levels (the platform for FQH physics) have been explored
\cite{Liu_Gromov_Papic_PRB2018_FQHquench,
Fremling_Haque_NJP2018_LLLdynamics}.  Dynamics in topological lattices (e.g., in Chern bands, or in the presence of a background lattice Berry curvature) have been studied both for wave packets \cite{Niu_arXiV2003,
Price_Cooper_PRA2012,
Roy_Grushin_Moessner_Haque_PRA2015, Dauphin_Massignan_Lewenstein_Lobo_SciPost2017} and for many-particle systems \cite{Killi_Paramekanti_PRA2012,
Killi_Trotzky_Paramekanti_PRA2012,
Goldman_Dalibard_Lewenstein_Zoller_Spielman_PNAS2013, Dauphin_Goldman_PRL2013,
Hauke_Lewenstein_Eckardt_tomography_PRL2014, Cooper_Bhaseen_quench_PRL2015,
Grushin_Roy_Haque_JSM2015, Yu_Haldanequench_phasevortices_PRA2017,
Dutta_Haldanequench_edgecurrents_PRB2018, Dong_Grushin_Motruk_Pollmann_PRL2018}.  Some of these theoretical investigations were motivated by developments with cold-atom experiments.  Indeed, real-time dynamical measurements are central to the study of topological matter/bands with cold atoms \cite{Aidelsburger_Abanin_Demler_Bloch_Zakphase_NatPhys2013,
  Esslingergroup_Haldanemodel_Nature2014, Aidelsberger_Cooper_Bloch_Goldman_NatPhys2015,
  Bloch_Schneider_interferometer_Science2015, Spielman_visualizing_edgestates_Science2015,
  Sengstock_Weitenberg_Berrycurvature_Floquet_2016experimental, Pan_Shanghai_experiment_PRL2018,
  Eckardt_Sengstock_Wittenberg_topologyfromdynamics_NatCom2019}.

We introduce the lattice Hamiltonians in Section \ref{sec:model}.  Following Ref.~\cite{Nandy_Nielsen_truncation_PRB2019}, we present and use the Hamiltonians in a form that distinguishes 1-body, 2-body and 3-body terms.  In Section \ref{sec:1p}, we present the spectrum and some dynamical phenomena of the one-particle sector, for square-shaped lattices.  The spectrum is found to be energetically arranged in order of decreasing distance from the center of the lattice: the lowest-energy eigenstates are concentrated spatially near the lattice center and the highest-energy states are localized at the lattice edges.  The propagation dynamics of an initially localized particle is found to be markedly chiral, while roughly maintaining its initial distance from the edge.  In Section \ref{sec:2p}, the two-particle system on square lattices is described: again we describe both spectra and dynamics.  The fermionic and hard-core bosonic systems are found to be qualitatively similar.  The qualitative features of the spectra and dynamics are explained as a combination of single-particle effects.  Finally, in Section \ref{sec:discussion}, we provide context by comparing with other lattice Hamiltonians which are topological or have chiral features.

\section{Lattice Hamiltonians}\label{sec:model}

In this section we define the lattice Hamiltonians and discuss some of their properties.

\subsection{Setup}

Our starting point is a family of lattice models, which were constructed in
Ref.~\cite{Tu_Nielsen_Cirac_Sierra_2014} using conformal field theory. The members of the family are labeled by the positive integer $q$, which is odd for fermions and even for hard-core bosons.  The inverse $1/q$ plays an analogous role as the filling fraction of Landau levels in the conventional quantum Hall effect.  If the lattice filling is set to be $1/q$, the ground state is known analytically to be a lattice analog of the $1/q$  Laughlin state. The Hamiltonians are given by
\begin{equation} \label{eq:master_Hamilt}
H=\sum_i\Lambda_i^\dag \Lambda_i, \quad
\Lambda_i = \sum_{j(\neq i)} w_{ij} [d_j-d_i(qn_j-1)],
\end{equation}
where $d_j$ is the annihilation operator acting on site $j$, and $n_j=d_j^\dag d_j$ is the particle number operator.  The particles are hardcore bosons (fermions) for $q$ even (odd). The coefficient is $w_{ij}=1/(z_i-z_j)$, where $z_j$ is the position of the $j$th lattice site written as a complex number $z_j=x_j+iy_j$.  The Hamiltonian is valid for general choices of $z_j$, but in this work we focus on the simplest case of a square lattice.  The Hamiltonian conserves the number of particles.

The Hamiltonian was designed to produce particular ground states at filling $1/q$.  However, having defined them on a particular lattice, we can consider any particle number sector, although the ground states for other fillings will not be Laughlin states.  In this work we focus on the smallest sectors: the 1-particle and 2-particle sectors.

\subsection{Decomposition into one-body, two-body and three-body parts}

We would like to split this Hamiltonian into a noninteracting part plus interactions.  By multiplying out the terms in the defining Hamiltonian, one obtains \cite{Nandy_Nielsen_truncation_PRB2019}
\begin{eqnarray} \label{eq:ham_5terms}
H &=& \sum_{i \ne j} C_1(i,j) \, d^{\dagger}_i d_j + \sum_{i \neq j} C_2(i,j) \, n_i n_j \nonumber \\
&+& \sum_{i \neq j \neq k} C_3(i,j,k) \, d^{\dagger}_i d_jn_k + \sum_{i \neq j \neq k} C_4(i,j,k) \, n_i n_jn_k \nonumber \\
&+& \sum_{i} C_5(i) \, n_i,
\end{eqnarray}
where the coefficients $C_i$ are given by
\begin{eqnarray}
&&C_1(i,j) = 2w^*_{ij}w_{ij} + \sum_{k (\neq i, \neq j)}(w^*_{ki}w_{kj} + w^*_{ji}w_{jk} + w^*_{ik}w_{ij}), \nonumber\\
&&C_2(i,j) = (q^2-2q)w^*_{ij}w_{ij} - q\sum_{k (\neq i,\neq j)}(w^*_{ij}w_{ik} + w^*_{ik}w_{ij}), \nonumber \\
&&C_3(i,j,k) = -q(w^*_{ji}w_{jk} + w^*_{ik}w_{ij}),  \label{coefddn} \nonumber \\
&&C_4(i,j,k) = q^2 w^*_{ik}w_{ij}, \label{coefnnn} \nonumber \\
&&C_5(i) = 2\sum_{j(\neq i)} w^*_{ij}w_{ij} + \sum_{j,k (\neq i)}w^*_{ik}w_{ij}. \label{coeffs}
\end{eqnarray}
The first and the fifth term of the Hamiltonian \eqref{eq:ham_5terms} involve only one particle, the second and the third term involve the interaction of two particles, and the fourth term involves three particles. If there is only one particle in the system, the Hamiltonian simplifies to
\begin{equation} \label{eq:ham1p}
 H_{1p} = \sum_{i \ne j} C_1(i,j) \, d^{\dagger}_i d_j + \sum_{i} C_5(i) \, n_i,
\end{equation}
The first ($C_1$) term is the single-particle hopping.  The hopping coefficients are complex and long-range.  Although hoppings at all distances are present, the magnitudes fall off slowly with distance.  The second ($C_5$) term is an on-site energy.  Note that the parameter $q$ does not
appear in either of these terms, consistent with the fact that quantum statistics does not play any role in the single-particle physics.

If there are two particles, the Hamiltonian simplifies to
\begin{eqnarray}
 H_{2p} &=& \sum_{i \ne j} C_1(i,j) \, d^{\dagger}_i d_j + \sum_{i \ne j} C_2(i,j) \, n_i n_j \nonumber \\
           &+& \sum_{i \ne j \ne k} C_3(i,j,k) \, d^{\dagger}_i d_jn_k + \sum_{i} C_5(i) \, n_i. \label{ham2p}
\end{eqnarray}
The new terms include both a density-density interaction term $C_2$ and a hopping term between any two sites that depends on the occupancy of other sites.  Again, these terms survive till arbitrary ranges.

If the total number of particles is $\geq 3$, all terms contribute and there is no simplification of the Hamiltonian.  The $C_4$ term is a three-body interaction term, which only appears when we have more than two particles in the system.  In this paper, we will focus on the one-particle and two-particle sectors, so that the $C_4$ term plays no role.

\begin{figure}
\begin{center}
\includegraphics[width =0.98\columnwidth]{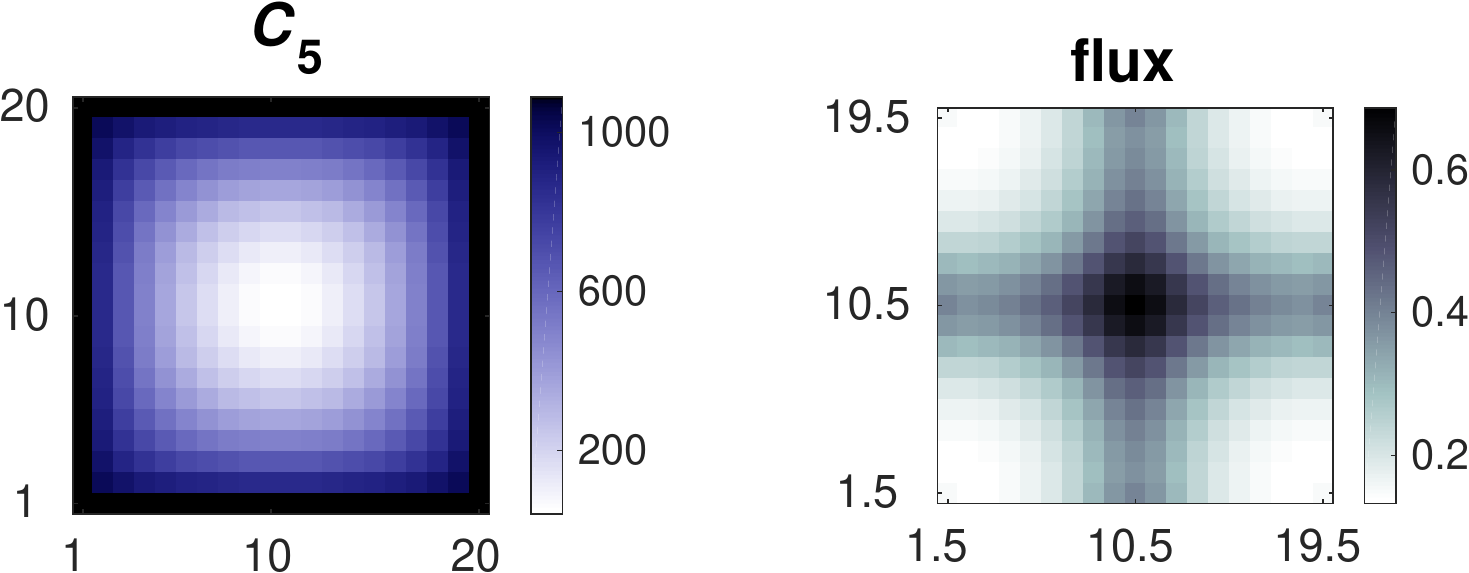}
\caption{Left: The potential $C_5$ for a $20 \times 20$ lattice.  The sites $i\equiv(i_x,i_y)$ are indexed by horizontal and vertical coordinates $i_x$ and $i_y$, which are each integers running from $1$ to $20$.  Right: Magnetic flux through each smallest plaquette for a $20 \times 20$ lattice.  The color at position $(i_x+0.5,i_y+0.5)$ in this plot represents the strength of the flux through the plaquette surrounded by the sites $(i_x,i_y)$, $(i_x+1,i_y)$, $(i_x,i_y+1)$, $(i_x+1,i_y+1)$.}
\label{C5}
\end{center}
\end{figure}

\subsection{Single-particle properties}

The noninteracting part $H_{1p}$ (Eq.\ \eqref{eq:ham1p}) of the Hamiltonian $H$, consists of a potential ($C_5$) and a complex hopping term ($C_1$).

In Figure \ref{C5} (left) we show the behavior of the potential $C_5$.  The potential increases rapidly as one moves from the center of the lattice towards the edges.  For square-shaped lattices, this increase is approximately quadratic in the horizontal and vertical directions, and somewhat more complicated in the diagonal directions.  From this perspective, the physics of the single-particle Hamiltonian is like that of a particle in a potential trap which is approximately harmonic.

When a particle hops around a closed loop, the state acquires an Aharonov-Bohm phase of $2\pi$ times the magnetic flux enclosed by the path. This allows us to compute the flux through each plaquette, by adding the phase factors of the nearest-neighbor hopping ($C_1$) terms surrounding that plaquette.  One can regard this as a `magnetic field'.  This quantity is shown in Figure \ref{C5} (right).  For a typical Landau level system, the magnetic field is uniform in real space, while for a Chern band, the Berry curvature has some structure in momentum space.  In contrast, in our model the flux strength has intricate spatial structure, being strongest in the center and weakest at the corners of the lattice.

The spatial variation observed in Fig.\ \ref{C5} originates from the shape of the edge of the lattice. This is so because $w_{ij}$ is only a function of the relative position of $z_i$ and $z_j$. This means that the local contributions to the sums appearing in the coefficients are the same for all sites in the bulk, while the variation comes from the nonlocal contributions, which are determined by the shape of the edge.

\subsection{Units and indexing}

In studies of quantum dynamics on lattice models, it is common practice to set the nearest-neighbor hopping strength to unity, so that time is measured in units of the inverse hopping strength.  Since we have many different hoppings, the nearest-neighbor hopping is not particularly special.  In the Hamiltonian \eqref{eq:ham_5terms}, for square-shaped lattices the nearest-neighbor hopping strength increases roughly linearly with the width/length of the lattice.

We find it convenient to use the form \eqref{eq:ham_5terms} for the Hamiltonian, without any rescalings.  This fixes the units of energy and time, which we do not explicitly specify.

In writing the Hamiltonians above, we have used $i$, $j$ to label sites.  Since we will be considering square (or rectangular) lattices of size $L_x\times L_y$, it will be convenient to sometimes use Cartesian coordinates for the site indices, e.g., $i\equiv (i_x,i_y)$.  The indices $i_x$ and $i_y$ will run from $1$ to $L_x$ and from $1$ to $L_y$ respectively.

\section{The one-particle sector --- spectrum and dynamics} \label{sec:1p}

In this section we examine the spectrum and some dynamical behaviors of the system with a single particle.  The evolution occurs according to the Hamiltonian \eqref{eq:ham1p}.  For definiteness we focus on a $20\times20$ square lattice but the results are very similar for other lattice sizes.

We first analyze the structure of the eigenvalues and eigenstates (Subsection \ref{subsec:1p_spectrum}), and then consider the dynamics when the single particle starts at various positions on the lattice (Subsection \ref{subsec:1p_dynamics}).

\subsection{Eigenvalues and eigenstates} \label{subsec:1p_spectrum}

\begin{figure}
\begin{center}
\includegraphics[width=0.99\columnwidth]{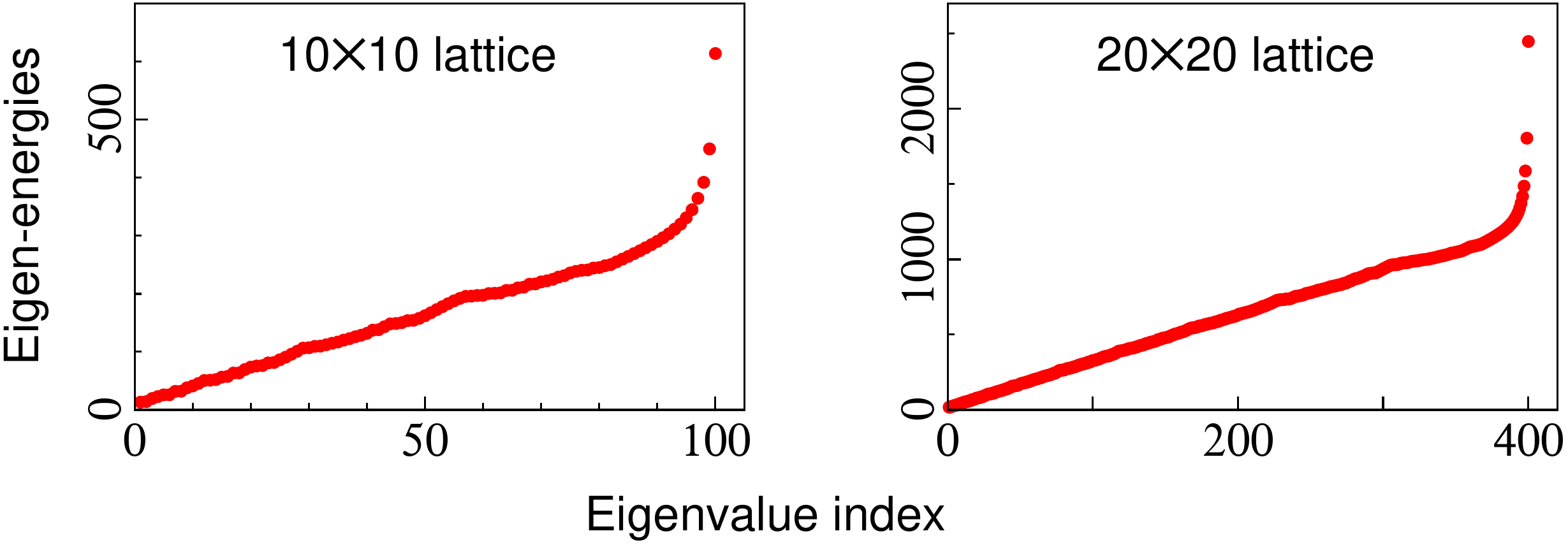}
\caption{Eigenenergies of the single-particle system, indexed in ascending order.}
\label{eigenvalues_1p}
\end{center}
\end{figure}

In Figure \ref{eigenvalues_1p} we show the eigenvalues of the square lattice subject to the single-particle Hamiltonian \eqref{eq:ham1p}, for sizes $10\times10$ and $20\times20$.  A prominent feature is that the spectrum is not clustered into separated bands.  In fact, when the eigenenergies are plotted against the index (ascending order of energies), the resulting plot is roughly linear over most of the spectrum, except for the last (highest-energy)  states which have rapidly increasing energy.
The absence of any band structure is a stark difference from the single-particle physics of, e.g., Chern-band lattices or Landau level physics.

\begin{figure}
\begin{center}
\includegraphics[width = 0.99\columnwidth]{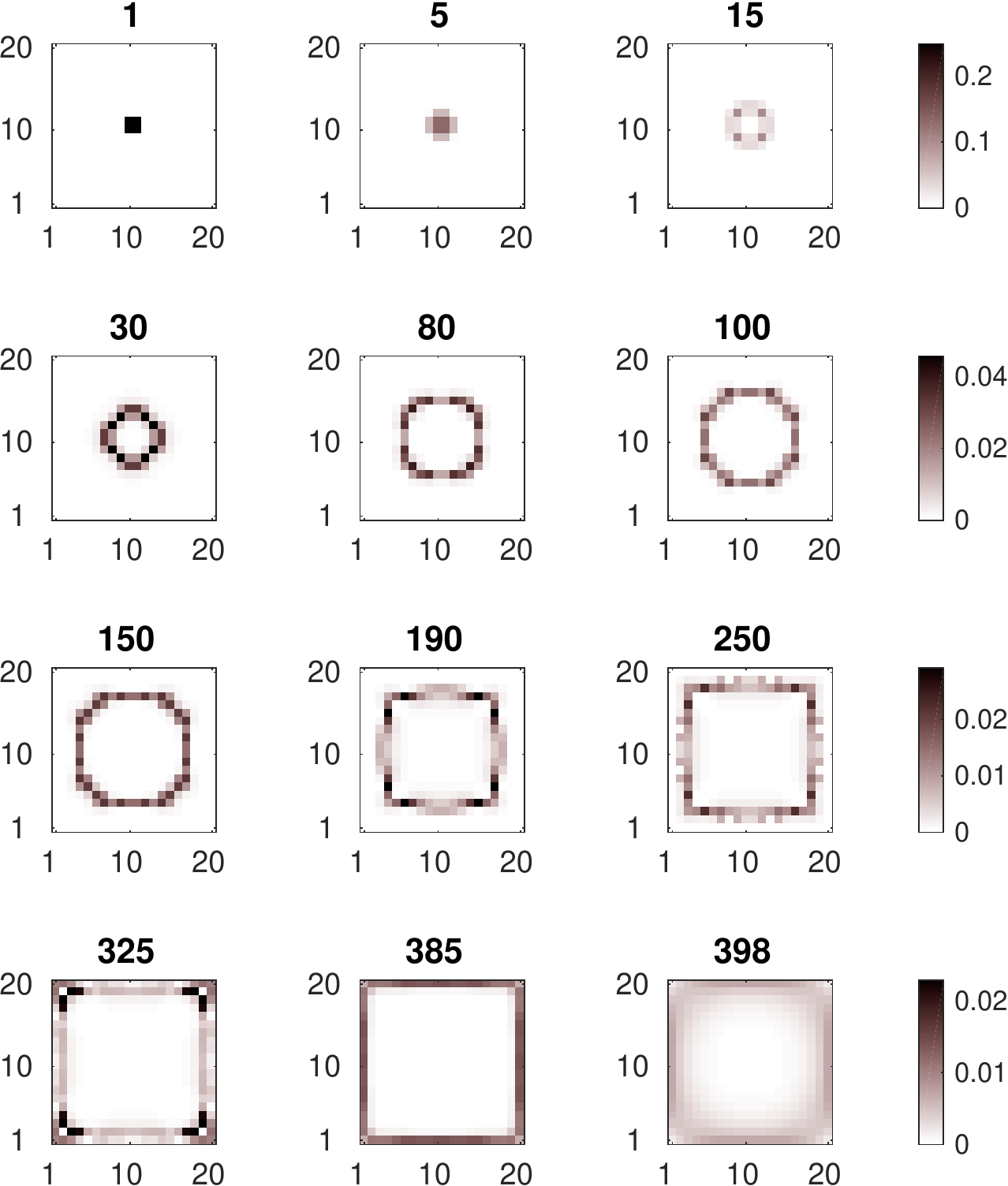}
\caption{Site occupancies $\langle E_{\alpha}|d_i^{\dagger}d_i|E_{\alpha}\rangle$ for different eigenstates $\ket{E_{\alpha}}$, where $\alpha$ indexes the eigenstates according to increasing energy.  The number above each plot indicates $\alpha$ (e.g.\ $\alpha=398$ for the last plot).  A different color scale has been used for each row, because in the higher eigenstates the weight is spread out among a larger number of sites, resulting in smaller nonzero site occupancies.
\label{density_eigen_1p}
}
\end{center}
\end{figure}

In Figure \ref{density_eigen_1p} we illustrate the spatial structure of the eigenstates, for the $20\times20$ lattice.  We have shown the density profiles of 12 of the eigenstates, starting from the ground state to one of the highest-energy eigenstates.  If the eigenstate corresponding to eigenvalue $E_\alpha$ is denoted by $\ket{E_\alpha}$, then the quantity displayed here is
\begin{equation}
 n_{i}(E_\alpha) = \bra{E_\alpha}  d_i^{\dagger}d_i \ket{E_\alpha} ,
\end{equation}
where $i$ is the site index.

The prominent feature of the density profiles is that each eigenstate is localized at a certain distance from the edges of the lattice.  The lower-energy eigenstates are localized far away from the edges, and the higher-energy eigenstates are localized close to the edges.  The ground state is very well-localized in the central four sites.  As we move up in energy, the eigenstates move out from the center toward the edges.  The $\approx40$ eigenstates with highest energies are mostly localized at the outermost layer of sites, although the degree of localization varies somewhat from eigenstate to eigenstate, as seen by comparing the $\alpha=385$ and $\alpha=398$ cases.  The eigenstates at intermediate energies are localized in regions which are generally of similar shape as the boundary, square in this case.  (However, the corners are significantly rounded for some of the eigenstates.)

The spatial structure of eigenstates is due to the potential $C_5$ (Figure \ref{C5} left).  The hopping magnitudes are relatively small compared to the values of the potential; hence each eigenstate consists of configurations with small variance of the $C_5$ term.  This is realized by sites at roughly fixed distance from the edges.  We have checked that, if the hopping terms ($C_1$) are replaced by nearest-neighbor hoppings of comparable magnitude, the eigenstates have similar spatial structure.

We have also checked that, for rectangular-shaped lattices ($L_x\neq L_y$), the overall features of the eigenvalues are very similar to those shown in Figure \ref{eigenvalues_1p}.  The eigenstates in such a lattice are localized in rectangular regions of similar aspect ratio as the lattice.

\subsection{Real-time dynamics} \label{subsec:1p_dynamics}

\begin{figure}
\begin{center}
\includegraphics[width=0.99\columnwidth]{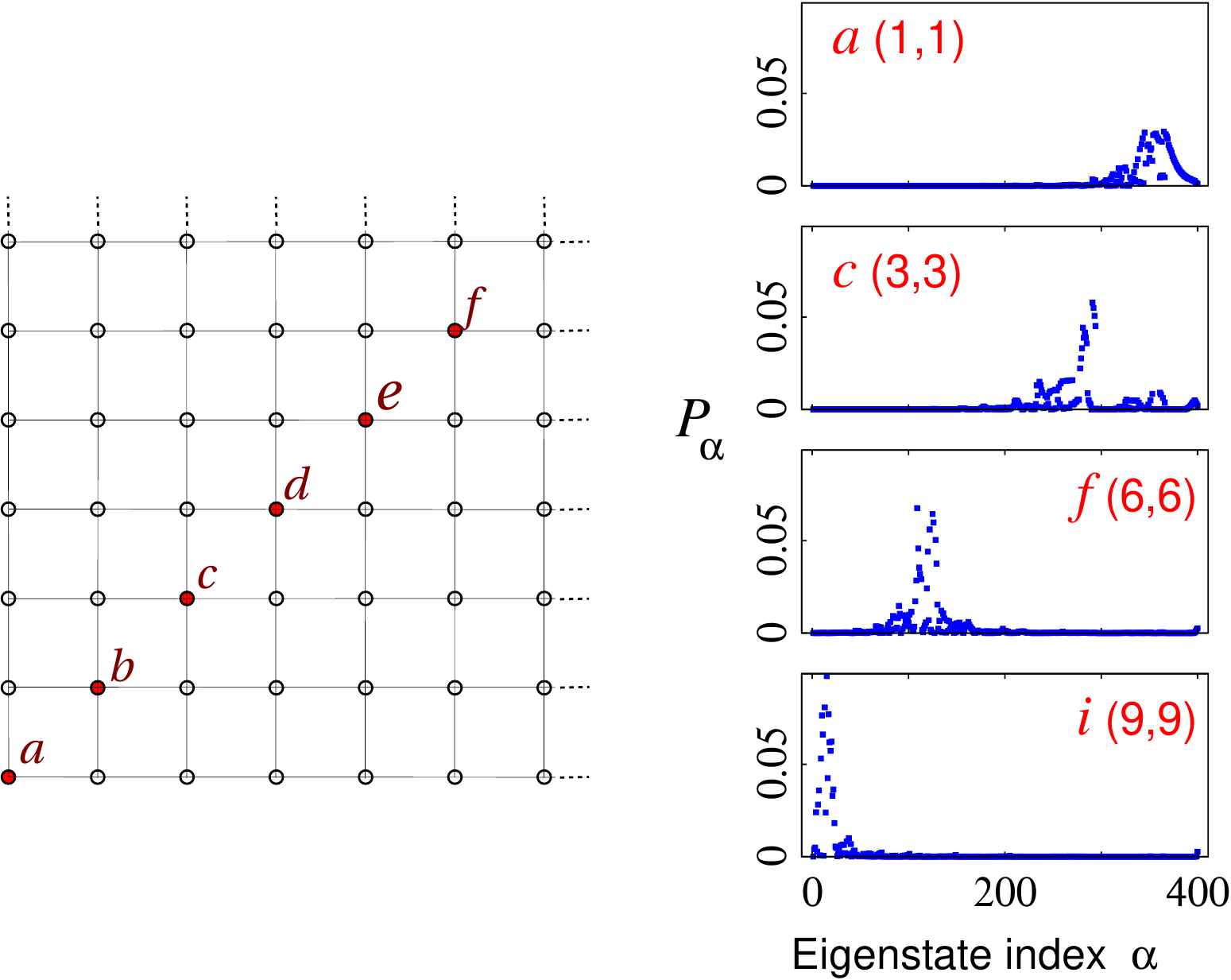}
\caption{Left: We choose the initial state \ket{\psi(0)} to be a state with a single particle at a particular site. The site could, e.g., be one of the sites $a$, $b$, $c$, etc. Right: Norm square $P_{\alpha}=\left| \innerprod{E_{\alpha}}{\psi(0)}\right|^2$ of the coefficients of the initial state in the basis of the energy eigenstates. For each plot, the legend gives the chosen initial position of the particle, and the lattice has $20 \times 20$ sites. One can notice from these plots that if we start the dynamics with the particle initially at the edge, then the particle is more likely to be found in the higher excited states. As we move towards the bulk, the contributing states have lower and lower energy.}
\label{prob_1p}
\end{center}
\end{figure}

\begin{figure*}
\begin{center}
\includegraphics[width = 0.82\textwidth]{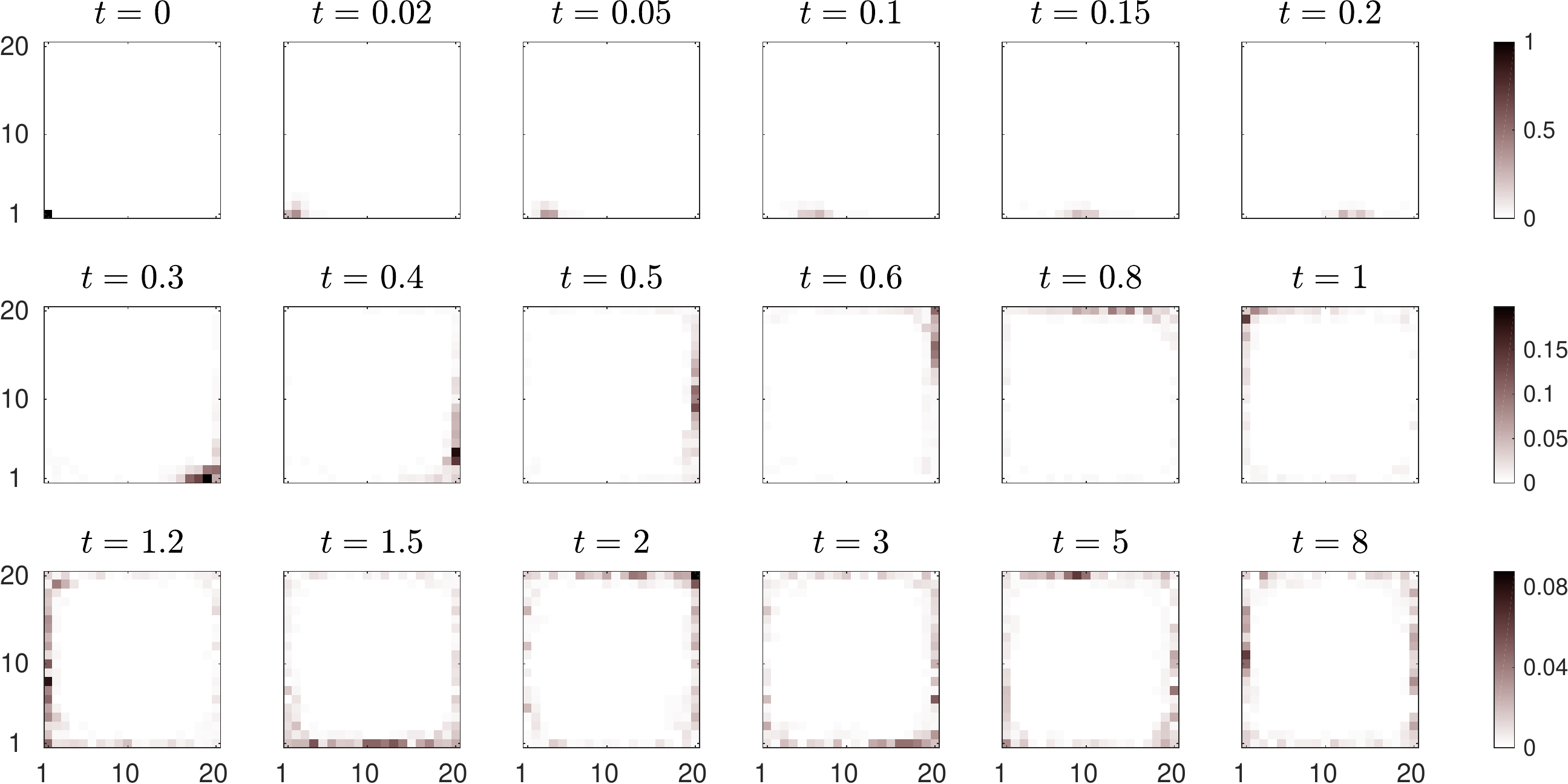}
\caption{Propagation dynamics of a single particle initially placed at the corner of a $20\times20$ lattice.  The dynamics is illustrated through a series of snapshots of the occupancy profiles, i.e., values of $\langle \psi(t)|d_i^{\dagger}d_i|\psi(t)\rangle$ at every site.  The particle is seen to execute chiral motion around the edge, in addition to dispersion along the edge on a longer timescale.  There is very little weight spreading into the bulk.} \label{dyn1p_ket_a}
\end{center}
\end{figure*}

\begin{figure*}
\begin{center}
\includegraphics[width = 0.82\textwidth]{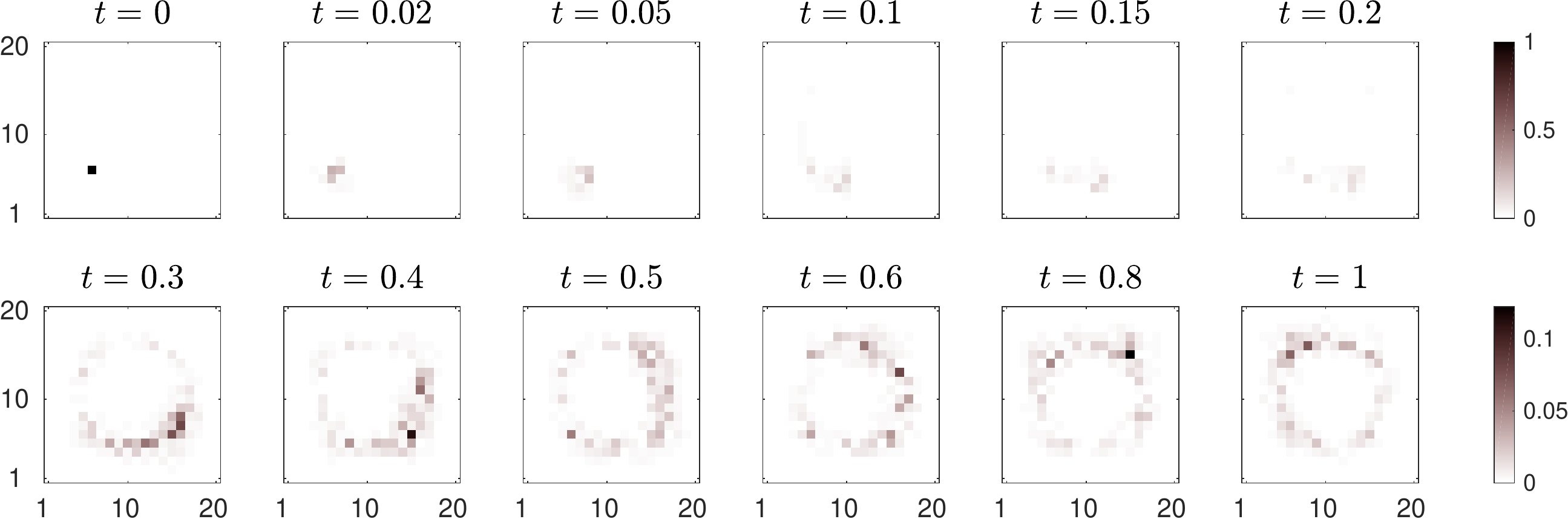}
\caption{Propagation dynamics of a single particle in a $20\times20$ lattice.  The particle is initially placed some distance away from the edge, at position $(6,6)$, referred to in the text as site $f$.  The particle is seen to move chirally along an orbit roughly equidistant from the edge.}
\label{dyn1p_ket_f}
\end{center}
\end{figure*}

We now consider real-time dynamics.  We will concentrate on initial states in which the particle is localized at a particular site.  As the initial position, we use sites at various distances from the edge, moving toward the center of the lattice.  For concreteness, we focus on the sites on the diagonal, starting from a corner site and then moving toward the center.  In Figure \ref{prob_1p} these sites are labeled as $a$, $b$, $c$,\ldots corresponding to the sites $(1,1)$, $(2,2)$, $(3,3)$,\ldots

The overlaps of an initial state with the system eigenstates affects the dynamics strongly, as can be seen from an expansion in the energy eigenbasis:
\begin{equation}
\ket{\psi(t)} = \sum_\alpha \innerprod{E_{\alpha}}{\psi(0)} e^{-iE_{\alpha}t} \ket{E_{\alpha}} .
\end{equation}
It is therefore useful to examine the overlaps.  In Figure \ref{prob_1p} (right), the magnitudes of overlap of some of the initial states with all the eigenstates are shown.  (Actually we show the squared magnitudes, $P_{\alpha} = \left| \innerprod{E_{\alpha}}{\psi(0)}\right|^2$.)  Since each initial state has the particle localized on a particular site, the overlap is simply the weight of the single-particle eigenstate on that site, and $P_{\alpha}$ is the probability of finding the particle at that site, when the system is in eigenstate $\ket{E_{\alpha}}$.  Knowing the spatial structure of the eigenstate, we expect that for an initial state with the particle localized at site $a$ (at the lattice edge), the overlap will mainly be with high-energy eigenstates.  From the same argument, one expects that as one moves the initial position toward the center of the lattice, the dominant overlaps move toward lower eigenenergies.  This pattern is seen clearly in Figure \ref{prob_1p}.

In Figure \ref{dyn1p_ket_a} we show the dynamics starting from position $a$ [corner site, $(1,1)$]. The dynamics is presented through snapshots of site occupancies at a series of time instants.  Since the corner site has appreciable overlap only with the highest-energy eigenstates, which are all localized near the edge of the lattice, we expect that the particle will stay confined to the edge of the lattice.  This is seen clearly in Figure \ref{dyn1p_ket_a}.  In addition, we see that the dynamics is chiral: the particle circles  counterclockwise around the edge of the lattice.  At the same time, there is increasing delocalization (dispersion) along the edge, so that, after $t\gtrsim1$, the wavefunction is spread throughout the edge regions.  There is however no significant delocalization perpendicular to the edge: the particle does not enter the bulk of the lattice.

We comment now on the timescale for the motion along the edge.  The particle seems to move by one site approximately in time $\approx0.015$.  For the $20\times20$ lattice, the nearest-neighbor hoppings along the edge bonds have magnitude between $\approx 48$ and $\approx 64$ (lower near the corners and higher near the edge centers).  Since $\hbar=1$, the hopping timescale can be expected to be the inverse of nearest-neighbor coupling strength, consistent with the observed $\approx0.015$.  Of course, this estimate does not take into account longer-range hoppings.

In Figure \ref{dyn1p_ket_f} we show the dynamics starting from position $f$ [site $(6,6)$], which is about halfway between the corner and the center of the lattice.  Since the eigenstates are each localized at different distances from the edge, this initial state will have overlap with those eigenstates which are at roughly this distance from the edge.  Thus, we expect the particle to remain confined in a region which is at this distance from the edge, i.e., to move along a roughly square-shaped ring-like region.  Another way of arguing this is to note that the quantum particle finds itself in a steep potential ($C_5$ term) on a two-dimensional lattice.  In such a situation, one finds that the particle is free to move perpendicular to the slope, and the only dynamics allowed in the slope direction would be small-amplitude Bloch oscillations, as studied, e.g., in \cite{Kolovsky_EPL2014, Roy_Grushin_Moessner_Haque_PRA2015}.

In Figure \ref{dyn1p_ket_f} we indeed see that the particle moves along a ring-shaped region and does not escape to regions closer to or farther from the edge.  In addition, we see the same chirality as we saw at the edge: the particle moves counterclockwise around this ring.  Eventually, the particle position also disperses along the ring-shaped orbit, so that the wavefunction is spread around the whole orbit.

When the particle is started at other distances from the edge, we see the same phenomenon: it moves chirally (counterclockwise) along an orbit that is roughly equidistant from the edge, and eventually there is dispersion that smears out the wavefunction over the complete orbit, i.e., the particle becomes delocalized parallel to the edge but remains localized perpendicular to the edge.

The localization behavior and the shapes of the orbits can be understood from the spatial structure of the eigenstates or, equivalently, from the shape of the on-site potential term $C_5$.  The chiral behavior of the motion is an additional intriguing feature.  The edge chirality (Figure \ref{dyn1p_ket_a}) is similar to the behavior in Chern band lattices, but the chirality in the bulk orbits (Figure \ref{dyn1p_ket_f}) is something peculiar to the present system.  We discuss this issue further in the concluding section.

\section{The two-particle sector} \label{sec:2p}

We now turn to the two-particle sector, described by the Hamiltonian \eqref{ham2p}.  In addition to the single-particle hopping ($C_1$) and potential ($C_5$) terms we have examined already, we now also have density-density interaction  ($C_2$) terms and correlated hopping ($C_3$) terms.

We consider both a system with two hard-core bosons ($q=2$, with $d_i$, $d_i^{\dagger}$ satisfying bosonic commutation relations), and a system with two spinless fermions ($q=3$, with the operators satisfying fermionic anti-commutation relations).  For the properties we will examine, the two cases will turn out to be very similar.

\begin{figure}
\begin{center}
\includegraphics[width = 0.9\columnwidth]{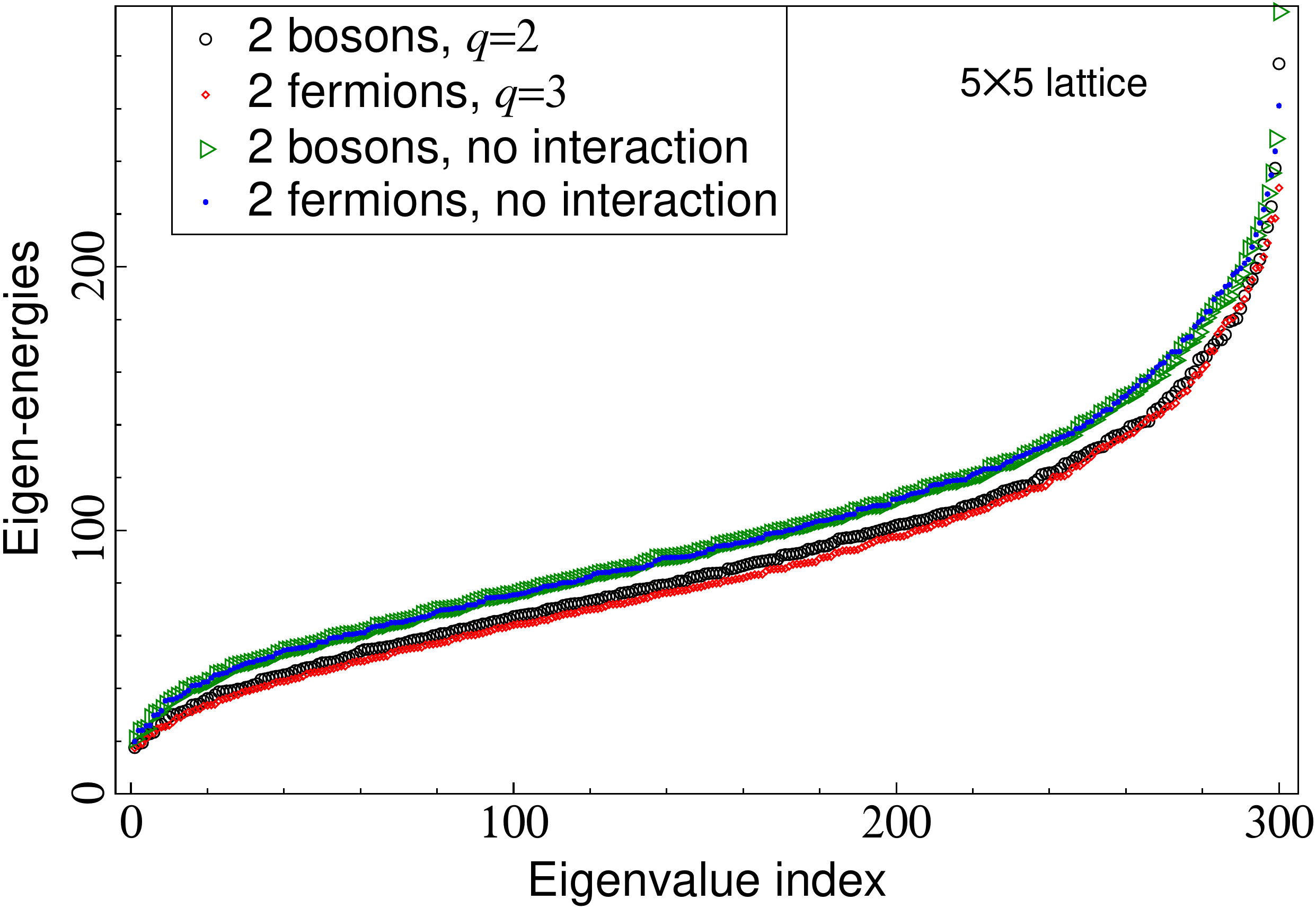}
\caption{Energy eigenvalues of the 2-particle systems, for a small ($5\times5$) square lattice.  The 2-fermion and 2-boson systems are seen to have very similar spectra (two lower datasets, barely distinguishable from each other).  Also shown (curves with slightly higher energies) are the spectra of 2-fermion and 2-boson systems in which the interaction terms ($C_2$ and $C_3$ terms) have been suppressed from the Hamiltonian.  These two datasets are also very close and hence barely distingushable from each other.
\label{eigenvalues_2p}
}
\end{center}
\end{figure}

\begin{figure}
\begin{center}
\includegraphics[width = 0.99\columnwidth]{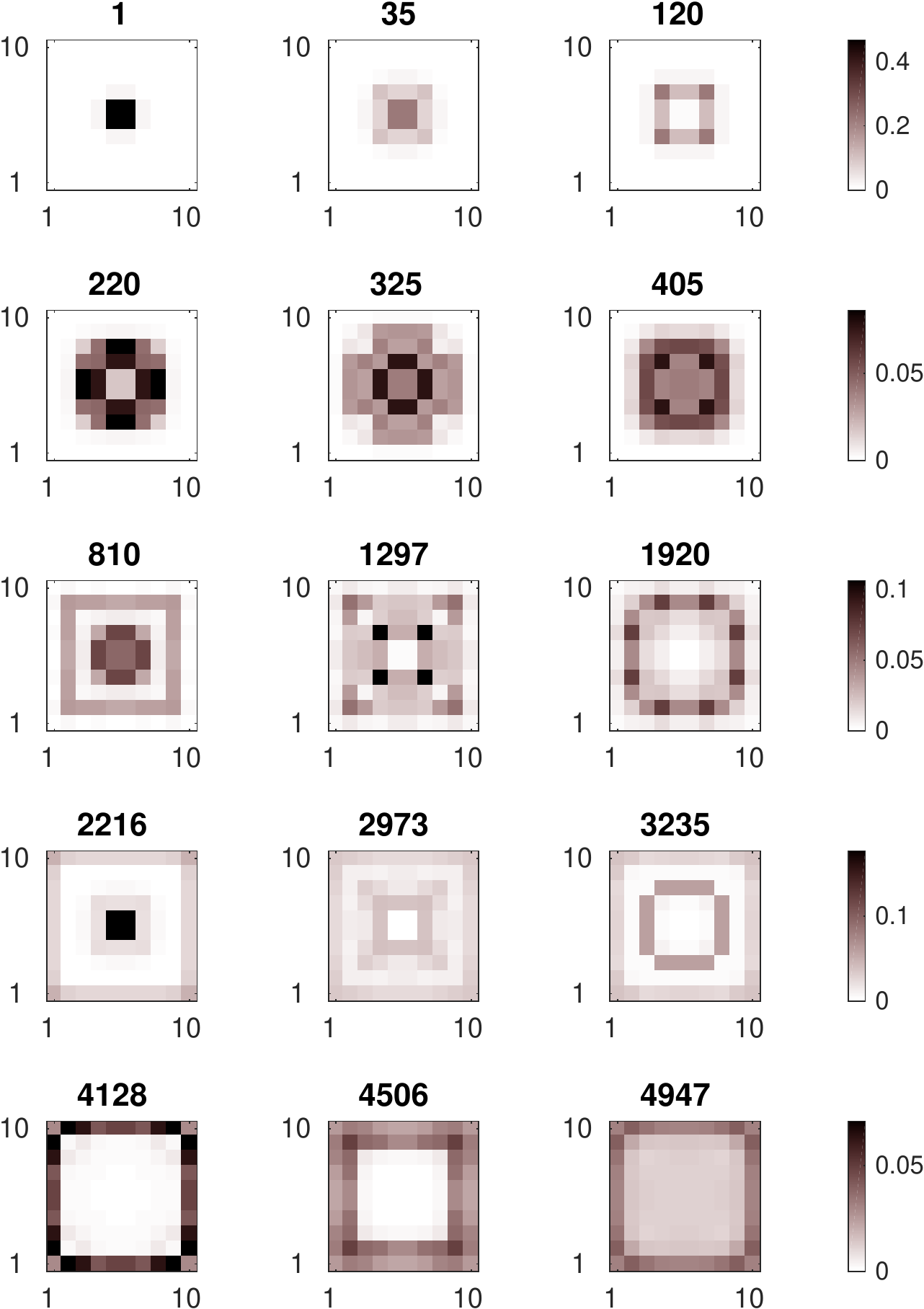}
\caption{Number density (site occupancy) profiles in different eigenstates of the two-fermion system. The eigenstate labels are indicated on top of the individual density plots.  A different color (density) scale has been used for each row, because in the higher eigenstates the weight is spread out among a larger number of sites, resulting in smaller nonzero site occupancies.  The two-boson case (not shown) shows the same overall features as one moves from low-energy to high-energy eigenstates.
\label{density_eigen_2p}}
\end{center}
\end{figure}

In Figure \ref{eigenvalues_2p}, we display the spectra of the 2-fermion and 2-boson systems for a square lattice.  A relatively small system ($5\times5$) is chosen for clarity; the features are qualitatively unchanged as the size is varied.  The 2-fermion and 2-boson spectra are seen to be very similar.

To evaluate the effect of the two-body terms, we also plot the spectra of the 2-fermion and 2-boson systems in which the $C_2$ and $C_3$ terms have been set to zero.  (In the fermionic case, this spectrum can be built by adding pairs of distinct single-particle eigenenergies.) The interacting systems are seen to have slightly smaller energies than the $C_2=C_3=0$ systems, but the difference is small.  We conclude that the effect of the two-body interaction/correlation terms is quite mild.

In Figure \ref{density_eigen_2p}, we display the density profiles for a selection of eigenstates for the two-fermion system on a $10\times10$ lattice.  The lowest-energy eigenstates have the two particles localized near the center of the lattice, while in the highest-energy eigenstates both particles are localized near the edges of the lattice.  At intermediate energies, the two particles are localized at various distances from the center.  In many cases, the density profile can be interpreted as two distinct single-particle eigenstates being occupied.  This qualitative picture is what one would expect of a system with non-interacting particles --- a visual inspection of the eigenstate density profiles does not show any obvious signature of the interaction/correlation terms.

The two-boson system (not shown) has the same qualitative features --- lowest-energy eigenstates have both particles near the center and highest-energy eigenstates have both near the edge.  Of course, the sequence of density profiles does not match in detail (e.g., the 810th eigenstate of the two-boson system does not have the same density profile as the 810th eigenstate of the two-fermion system).

\begin{figure*}
\begin{center}
\includegraphics[width=0.82\textwidth]{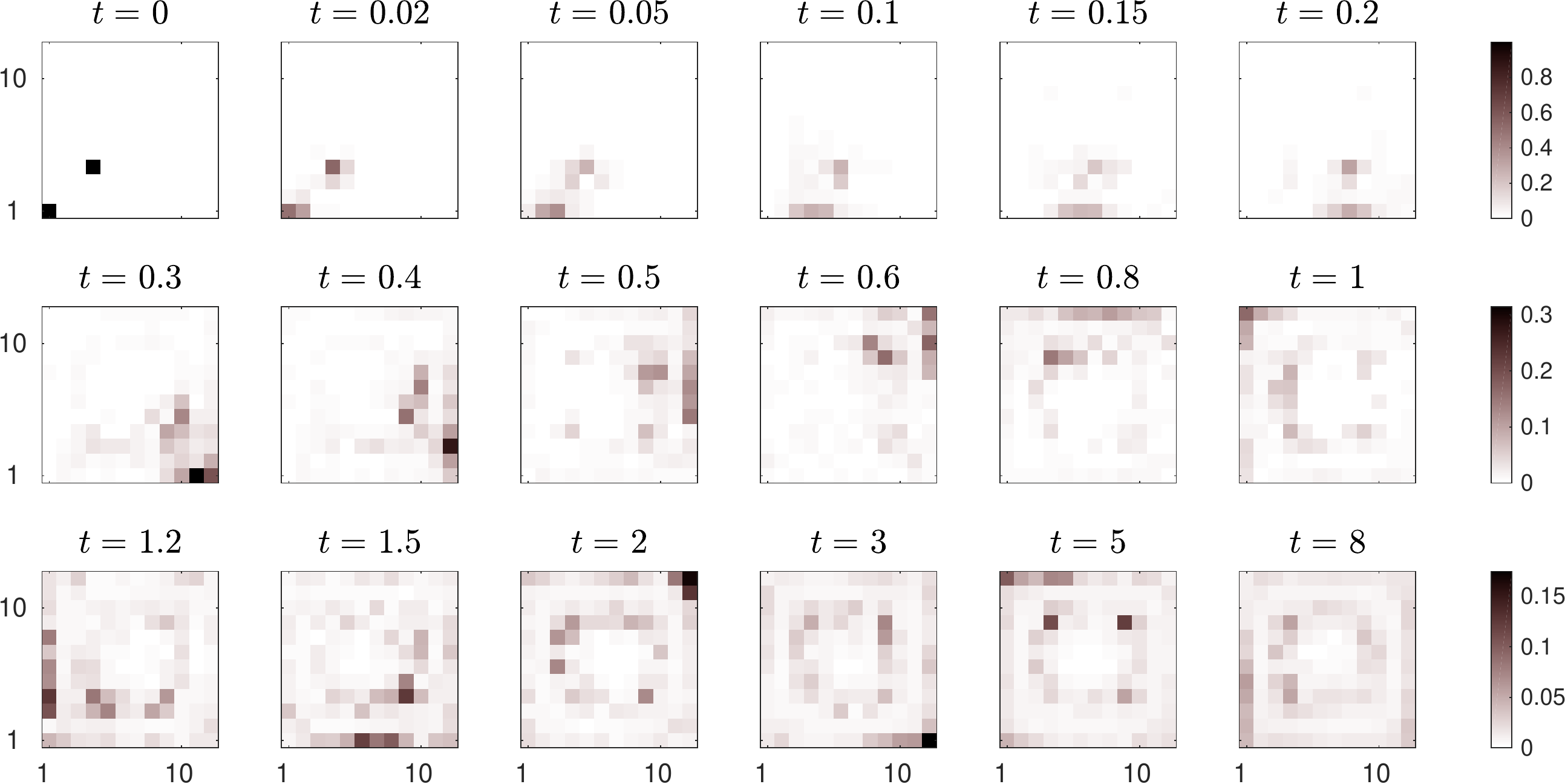}
\caption{Dynamics of the two-boson system on a $12\times12$ lattice.  In the initial state, the two bosons are at the positions $a$ and $d$, i.e., at the corner site $(1,1)$ and at the site $(4,4)$. }
\label{dyn2p_ket_a_d}
\end{center}
\end{figure*}

This structure of the eigenstates is reflected in the dynamics.  When both particles are started near the center, they remain near the center, while if both particles are started at/near the edge, they each perform chiral dynamics at/near the edge.  There is no qualitative difference seen between fermionic and bosonic cases.

In Figure \ref{dyn2p_ket_a_d}, we show the dynamics following from placing two particles at different distances from the edge.  Each particle follows roughly the orbit that it would have followed in the single-particle case, and eventually the wavefunction spreads in the regions covered by these two orbits.

\section{Concluding discussion} \label{sec:discussion}

In this work, we have examined the one-particle and two-particle sectors of lattice models which, at appropriate fillings, have Laughlin states as exact ground states.  We have focused on square-shaped lattices, but expect the qualitative results to be valid for other lattice shapes.

\paragraph*{No Chern bands ---}
We find that the single-particle spectrum is not arranged in bands.  This means that the mechanism for obtaining fractional quantum Hall states in the present class of models is strikingly different from that in fractional Chern insulators \cite{NeupertChamonMudry_FCI_PRL2011,RegnaultBernevig_PRX2011} or in Hofstadter lattices \cite{SorensenDemlerLukin_PRL2005,HafeziSorensenDemlerLukin_PRA2007}.  In those cases, the single-particle system is composed of bands with definite Chern number, and adding interactions to partially filled Chern bands is expected to create FQH ground states in a manner analogous to FQH states observed in fractionally filled Landau levels in the continuum.  In Hamiltonians of the type \eqref{eq:master_Hamilt} derived from CFT, our finding indicates that the FQH phenomenon is not reliant on Chern band physics in the single-particle sector.  The present Hamiltonians are not easy to adapt to periodic boundary conditions, which means Chern numbers would anyway be tricky to define.

\paragraph*{Ring-shaped eigenstates and orbits ---}
The single-particle spectrum is arranged energetically according to eigenstates being progressively closer to the edge of the lattice.  This is not surprising given that the single-particle potential term ($C_5$ term) in the Hamiltonian acts as a strong `trapping' potential.  Single particle states in trapping potentials have been previously explored in, e.g., \cite{Kolovsky_Grusdt_Fleischhauer_PRA2014, Grushin_Roy_Haque_JSM2015}.  Similar radially localized eigenstates were found in those cases.

Given this spatial structure of the single-particle eigenstates, it is unsurprising that the spreading/propagation dynamics of an initially localized particle occurs around such a square/ring-shaped region.  The dynamics in Refs.~\cite{Kolovsky_Grusdt_Fleischhauer_PRA2014,Roy_Grushin_Moessner_Haque_PRA2015} is somewhat comparable.  One might also expect to see some Bloch oscillation in the direction perpendicular to the square/ring-shaped orbit, but the amplitude of such oscillations are too small to be visible in our occupancy snapshots.

\paragraph*{Chiral motion ---}
While the radially localized orbits could be expected, the fact that the trajectories are \emph{chiral} is a rather surprising aspect.  It is interesting to compare with other systems.  In Chern band systems with open boundary conditions, the only single-particle eigenstates which have overlap with the edge are the edge states which connect two bands.  These edge states have clear chirality: when the eigenenergies are plotted as a function of momentum, the edge states appear as lines with a definite slope in one direction.  Thus, if a single particle is placed at the edge of an open-boundary Chern band system (e.g., the Haldane lattice \cite{Haldane_PRL1988}), one sees chiral motion due to the chirality of the edge states.  This type of chiral motion has been demonstrated experimentally in a photonic system \cite{Szameit_photonic_topological_expt_Nature2013}.  However, a particle placed in the \emph{bulk} of a Chern-band lattice does not have chiral motion, and should disperse in all directions.  Thus our observed chiral motion, which occurs both along the edge and in the bulk, cannot be regarded as analogous to the Chern-band lattice case.  (The dynamics in the bulk that we show in Figure \ref{dyn1p_ket_f} is utterly different from what would be observed in a Chern-band lattice.)
In the system studied in the present paper, we cannot identify chirality by plotting energy versus momentum, as momentum is difficult to define.

Other than the edges of Chern-band lattices, single-particle dynamics on lattice systems do not generally show chiral motion.  For example, if one considers a tight-binding ring threaded by a magnetic flux (through Peierls fluxes on the hopping amplitudes), a particle released at one site will propagate in both directions.

On a Hofstadter lattice (say open-boundary square lattice with uniform flux per plaquette), a particle placed in the bulk spreads out in all directions. A particle placed at the edge does spread out along the edge and avoids spreading into the bulk (as in our case), but this motion is not chiral --- the particle spreads out in \emph{both} directions.  In such a case, the particle has overlap with bands with Chern numbers of both signs.  To examine the effect of the different Chern bands, we have done the following numerical experiment: Starting from the particle on the edge, one converts into the basis of energy eigenstates and projects away all contributions from the upper half of the energy spectrum, keeping only states with one particular sign of the Chern number.  (Of course, the initial particle wavefunction is no longer localized perfectly on a single site, but is spread over several sites.)  In this case, performing the real-time dynamics, we have found that the particle moves chirally around the edge.  Thus, even with the chiral Hofstadter Hamiltonian, one needs to construct special initial states to obtain chiral motion.

In Ref.~\cite{Kolovsky_Grusdt_Fleischhauer_PRA2014}, the dynamics of a wavepacket has been reported for a Hofstadter lattice subject to a harmonic trapping potential.  In that case, the wavepacket spreads out non-chirally across a ring-shaped region, or in extreme cases remains localized.

In addition to lattice systems, we could also compare with the dynamics of a single particle in the continuum, subject to both a magnetic field and a potential, and confined to the lowest Landau level.  An initially localized quantum particle in this situation would presumably follow dynamics broadly similar to what is observed in the present case --- chiral dynamics along an equipotential orbit, together with some dispersion of the wavepacket.

\paragraph*{Two-particle sector ---}
For the features of the two-particle sector that we have examined, the interaction or quantum statistics seems to play little role.  The particles mostly behave as if they were independent particles, and there is little qualitative difference between fermionic and bosonic cases.  Somewhat disappointingly, the propagation dynamics does not show any obvious signatures of the rather unusual $C_3$ term, which has the form of `assisted' hopping.

\paragraph*{Outlook ---}  In this work we have studied the simplest dynamical aspects of the CFT-inspired Hamiltonians introduced in Refs.~\cite{Tu_Nielsen_Cirac_Sierra_2014}, and found some surprising aspects already in the single-particle and two-particle sectors.  One might imagine new dynamical phenomena at finite filling, but numerical study of dynamics for appreciable sizes becomes challenging at finite fillings.  More generally, we expect that propagation and spreading dynamics in topological systems, in particular in the presence of edges and spatial inhomogeneities as in the present case, will give rise to further insights into the structure of different classes of Hamiltonians supporting topological matter.

\begin{acknowledgments}
DKN would like to thank the Max Planck Institute for the Physics of Complex Systems for hospitality during visits to the institute. This work has been supported by the Villum Foundation and by the (Polish) National Science Center Grant No.\ 2016/22/E/ST2/00555.
\end{acknowledgments}


\begin{thebibliography}{42}%
\makeatletter
\providecommand \@ifxundefined [1]{%
 \@ifx{#1\undefined}
}%
\providecommand \@ifnum [1]{%
 \ifnum #1\expandafter \@firstoftwo
 \else \expandafter \@secondoftwo
 \fi
}%
\providecommand \@ifx [1]{%
 \ifx #1\expandafter \@firstoftwo
 \else \expandafter \@secondoftwo
 \fi
}%
\providecommand \natexlab [1]{#1}%
\providecommand \enquote  [1]{``#1''}%
\providecommand \bibnamefont  [1]{#1}%
\providecommand \bibfnamefont [1]{#1}%
\providecommand \citenamefont [1]{#1}%
\providecommand \href@noop [0]{\@secondoftwo}%
\providecommand \href [0]{\begingroup \@sanitize@url \@href}%
\providecommand \@href[1]{\@@startlink{#1}\@@href}%
\providecommand \@@href[1]{\endgroup#1\@@endlink}%
\providecommand \@sanitize@url [0]{\catcode `\\12\catcode `\$12\catcode
  `\&12\catcode `\#12\catcode `\^12\catcode `\_12\catcode `\%12\relax}%
\providecommand \@@startlink[1]{}%
\providecommand \@@endlink[0]{}%
\providecommand \url  [0]{\begingroup\@sanitize@url \@url }%
\providecommand \@url [1]{\endgroup\@href {#1}{\urlprefix }}%
\providecommand \urlprefix  [0]{URL }%
\providecommand \Eprint [0]{\href }%
\providecommand \doibase [0]{https://doi.org/}%
\providecommand \selectlanguage [0]{\@gobble}%
\providecommand \bibinfo  [0]{\@secondoftwo}%
\providecommand \bibfield  [0]{\@secondoftwo}%
\providecommand \translation [1]{[#1]}%
\providecommand \BibitemOpen [0]{}%
\providecommand \bibitemStop [0]{}%
\providecommand \bibitemNoStop [0]{.\EOS\space}%
\providecommand \EOS [0]{\spacefactor3000\relax}%
\providecommand \BibitemShut  [1]{\csname bibitem#1\endcsname}%
\let\auto@bib@innerbib\@empty
\bibitem [{\citenamefont {Laughlin}(1983)}]{Laughlin_PRL1983}%
  \BibitemOpen
  \bibfield  {author} {\bibinfo {author} {\bibfnamefont {R.~B.}\ \bibnamefont
  {Laughlin}},\ }\bibfield  {title} {\bibinfo {title} {Anomalous quantum {H}all
  effect: An incompressible quantum fluid with fractionally charged
  excitations},\ }\href {https://doi.org/10.1103/PhysRevLett.50.1395}
  {\bibfield  {journal} {\bibinfo  {journal} {Phys. Rev. Lett.}\ }\textbf
  {\bibinfo {volume} {50}},\ \bibinfo {pages} {1395} (\bibinfo {year}
  {1983})}\BibitemShut {NoStop}%
\bibitem [{\citenamefont {Wen}(1989)}]{Wen_PRB1989}%
  \BibitemOpen
  \bibfield  {author} {\bibinfo {author} {\bibfnamefont {X.~G.}\ \bibnamefont
  {Wen}},\ }\bibfield  {title} {\bibinfo {title} {Vacuum degeneracy of chiral
  spin states in compactified space},\ }\href
  {https://doi.org/10.1103/PhysRevB.40.7387} {\bibfield  {journal} {\bibinfo
  {journal} {Phys. Rev. B}\ }\textbf {\bibinfo {volume} {40}},\ \bibinfo
  {pages} {7387} (\bibinfo {year} {1989})}\BibitemShut {NoStop}%
\bibitem [{\citenamefont {Wen}(1990)}]{Wen_IJMP1990}%
  \BibitemOpen
  \bibfield  {author} {\bibinfo {author} {\bibfnamefont {X.~G.}\ \bibnamefont
  {Wen}},\ }\bibfield  {title} {\bibinfo {title} {Topological orders in rigid
  states},\ }\href {https://doi.org/10.1142/S0217979290000139} {\bibfield
  {journal} {\bibinfo  {journal} {International Journal of Modern Physics B}\
  }\textbf {\bibinfo {volume} {04}},\ \bibinfo {pages} {239} (\bibinfo {year}
  {1990})}\BibitemShut {NoStop}%
\bibitem [{\citenamefont {Wen}\ and\ \citenamefont
  {Niu}(1990)}]{WenNiu_PRB1990}%
  \BibitemOpen
  \bibfield  {author} {\bibinfo {author} {\bibfnamefont {X.~G.}\ \bibnamefont
  {Wen}}\ and\ \bibinfo {author} {\bibfnamefont {Q.}~\bibnamefont {Niu}},\
  }\bibfield  {title} {\bibinfo {title} {Ground-state degeneracy of the
  fractional quantum {H}all states in the presence of a random potential and on
  high-genus {R}iemann surfaces},\ }\href
  {https://doi.org/10.1103/PhysRevB.41.9377} {\bibfield  {journal} {\bibinfo
  {journal} {Phys. Rev. B}\ }\textbf {\bibinfo {volume} {41}},\ \bibinfo
  {pages} {9377} (\bibinfo {year} {1990})}\BibitemShut {NoStop}%
\bibitem [{\citenamefont {Kalmeyer}\ and\ \citenamefont
  {Laughlin}(1987)}]{KalmeyerLaughlin_PRL1987}%
  \BibitemOpen
  \bibfield  {author} {\bibinfo {author} {\bibfnamefont {V.}~\bibnamefont
  {Kalmeyer}}\ and\ \bibinfo {author} {\bibfnamefont {R.~B.}\ \bibnamefont
  {Laughlin}},\ }\bibfield  {title} {\bibinfo {title} {Equivalence of the
  resonating-valence-bond and fractional quantum {H}all states},\ }\href
  {https://doi.org/10.1103/PhysRevLett.59.2095} {\bibfield  {journal} {\bibinfo
   {journal} {Phys. Rev. Lett.}\ }\textbf {\bibinfo {volume} {59}},\ \bibinfo
  {pages} {2095} (\bibinfo {year} {1987})}\BibitemShut {NoStop}%
\bibitem [{\citenamefont {Schroeter}\ \emph {et~al.}(2007)\citenamefont
  {Schroeter}, \citenamefont {Kapit}, \citenamefont {Thomale},\ and\
  \citenamefont {Greiter}}]{Thomale_Greiter_chiralspinliquid_PRL2007}%
  \BibitemOpen
  \bibfield  {author} {\bibinfo {author} {\bibfnamefont {D.~F.}\ \bibnamefont
  {Schroeter}}, \bibinfo {author} {\bibfnamefont {E.}~\bibnamefont {Kapit}},
  \bibinfo {author} {\bibfnamefont {R.}~\bibnamefont {Thomale}},\ and\ \bibinfo
  {author} {\bibfnamefont {M.}~\bibnamefont {Greiter}},\ }\bibfield  {title}
  {\bibinfo {title} {Spin {H}amiltonian for which the chiral spin liquid is the
  exact ground state},\ }\href {https://doi.org/10.1103/PhysRevLett.99.097202}
  {\bibfield  {journal} {\bibinfo  {journal} {Phys. Rev. Lett.}\ }\textbf
  {\bibinfo {volume} {99}},\ \bibinfo {pages} {097202} (\bibinfo {year}
  {2007})}\BibitemShut {NoStop}%
\bibitem [{\citenamefont {Yao}\ and\ \citenamefont
  {Kivelson}(2007)}]{Yao_Kivelson_chiralspinliquid_PRL2007}%
  \BibitemOpen
  \bibfield  {author} {\bibinfo {author} {\bibfnamefont {H.}~\bibnamefont
  {Yao}}\ and\ \bibinfo {author} {\bibfnamefont {S.~A.}\ \bibnamefont
  {Kivelson}},\ }\bibfield  {title} {\bibinfo {title} {Exact chiral spin liquid
  with non-{A}belian anyons},\ }\href
  {https://doi.org/10.1103/PhysRevLett.99.247203} {\bibfield  {journal}
  {\bibinfo  {journal} {Phys. Rev. Lett.}\ }\textbf {\bibinfo {volume} {99}},\
  \bibinfo {pages} {247203} (\bibinfo {year} {2007})}\BibitemShut {NoStop}%
\bibitem [{\citenamefont {S\o{}rensen}\ \emph {et~al.}(2005)\citenamefont
  {S\o{}rensen}, \citenamefont {Demler},\ and\ \citenamefont
  {Lukin}}]{SorensenDemlerLukin_PRL2005}%
  \BibitemOpen
  \bibfield  {author} {\bibinfo {author} {\bibfnamefont {A.~S.}\ \bibnamefont
  {S\o{}rensen}}, \bibinfo {author} {\bibfnamefont {E.}~\bibnamefont
  {Demler}},\ and\ \bibinfo {author} {\bibfnamefont {M.~D.}\ \bibnamefont
  {Lukin}},\ }\bibfield  {title} {\bibinfo {title} {Fractional quantum {H}all
  states of atoms in optical lattices},\ }\href
  {https://doi.org/10.1103/PhysRevLett.94.086803} {\bibfield  {journal}
  {\bibinfo  {journal} {Phys. Rev. Lett.}\ }\textbf {\bibinfo {volume} {94}},\
  \bibinfo {pages} {086803} (\bibinfo {year} {2005})}\BibitemShut {NoStop}%
\bibitem [{\citenamefont {Hafezi}\ \emph {et~al.}(2007)\citenamefont {Hafezi},
  \citenamefont {S\o{}rensen}, \citenamefont {Demler},\ and\ \citenamefont
  {Lukin}}]{HafeziSorensenDemlerLukin_PRA2007}%
  \BibitemOpen
  \bibfield  {author} {\bibinfo {author} {\bibfnamefont {M.}~\bibnamefont
  {Hafezi}}, \bibinfo {author} {\bibfnamefont {A.~S.}\ \bibnamefont
  {S\o{}rensen}}, \bibinfo {author} {\bibfnamefont {E.}~\bibnamefont
  {Demler}},\ and\ \bibinfo {author} {\bibfnamefont {M.~D.}\ \bibnamefont
  {Lukin}},\ }\bibfield  {title} {\bibinfo {title} {Fractional quantum {H}all
  effect in optical lattices},\ }\href
  {https://doi.org/10.1103/PhysRevA.76.023613} {\bibfield  {journal} {\bibinfo
  {journal} {Phys. Rev. A}\ }\textbf {\bibinfo {volume} {76}},\ \bibinfo
  {pages} {023613} (\bibinfo {year} {2007})}\BibitemShut {NoStop}%
\bibitem [{\citenamefont {Neupert}\ \emph {et~al.}(2011)\citenamefont
  {Neupert}, \citenamefont {Santos}, \citenamefont {Chamon},\ and\
  \citenamefont {Mudry}}]{NeupertChamonMudry_FCI_PRL2011}%
  \BibitemOpen
  \bibfield  {author} {\bibinfo {author} {\bibfnamefont {T.}~\bibnamefont
  {Neupert}}, \bibinfo {author} {\bibfnamefont {L.}~\bibnamefont {Santos}},
  \bibinfo {author} {\bibfnamefont {C.}~\bibnamefont {Chamon}},\ and\ \bibinfo
  {author} {\bibfnamefont {C.}~\bibnamefont {Mudry}},\ }\bibfield  {title}
  {\bibinfo {title} {Fractional quantum {H}all states at zero magnetic field},\
  }\href {https://doi.org/10.1103/PhysRevLett.106.236804} {\bibfield  {journal}
  {\bibinfo  {journal} {Phys. Rev. Lett.}\ }\textbf {\bibinfo {volume} {106}},\
  \bibinfo {pages} {236804} (\bibinfo {year} {2011})}\BibitemShut {NoStop}%
\bibitem [{\citenamefont {Regnault}\ and\ \citenamefont
  {Bernevig}(2011)}]{RegnaultBernevig_PRX2011}%
  \BibitemOpen
  \bibfield  {author} {\bibinfo {author} {\bibfnamefont {N.}~\bibnamefont
  {Regnault}}\ and\ \bibinfo {author} {\bibfnamefont {B.~A.}\ \bibnamefont
  {Bernevig}},\ }\bibfield  {title} {\bibinfo {title} {Fractional chern
  insulator},\ }\href {https://doi.org/10.1103/PhysRevX.1.021014} {\bibfield
  {journal} {\bibinfo  {journal} {Phys. Rev. X}\ }\textbf {\bibinfo {volume}
  {1}},\ \bibinfo {pages} {021014} (\bibinfo {year} {2011})}\BibitemShut
  {NoStop}%
\bibitem [{\citenamefont {Tu}\ \emph {et~al.}(2014)\citenamefont {Tu},
  \citenamefont {Nielsen}, \citenamefont {Cirac},\ and\ \citenamefont
  {Sierra}}]{Tu_Nielsen_Cirac_Sierra_2014}%
  \BibitemOpen
  \bibfield  {author} {\bibinfo {author} {\bibfnamefont {H.-H.}\ \bibnamefont
  {Tu}}, \bibinfo {author} {\bibfnamefont {A.~E.~B.}\ \bibnamefont {Nielsen}},
  \bibinfo {author} {\bibfnamefont {J.~I.}\ \bibnamefont {Cirac}},\ and\
  \bibinfo {author} {\bibfnamefont {G.}~\bibnamefont {Sierra}},\ }\bibfield
  {title} {\bibinfo {title} {Lattice {L}aughlin states of bosons and fermions
  at filling fractions $1/q$},\ }\href
  {https://doi.org/10.1088/1367-2630/16/3/033025} {\bibfield  {journal}
  {\bibinfo  {journal} {New Journal of Physics}\ }\textbf {\bibinfo {volume}
  {16}},\ \bibinfo {pages} {033025} (\bibinfo {year} {2014})}\BibitemShut
  {NoStop}%
\bibitem [{\citenamefont {Glasser}\ \emph {et~al.}(2016)\citenamefont
  {Glasser}, \citenamefont {Cirac}, \citenamefont {Sierra},\ and\ \citenamefont
  {Nielsen}}]{Glasser_Cirac_Sierra_Nielsen_2016}%
  \BibitemOpen
  \bibfield  {author} {\bibinfo {author} {\bibfnamefont {I.}~\bibnamefont
  {Glasser}}, \bibinfo {author} {\bibfnamefont {J.~I.}\ \bibnamefont {Cirac}},
  \bibinfo {author} {\bibfnamefont {G.}~\bibnamefont {Sierra}},\ and\ \bibinfo
  {author} {\bibfnamefont {A.~E.~B.}\ \bibnamefont {Nielsen}},\ }\bibfield
  {title} {\bibinfo {title} {Lattice effects on {L}aughlin wave functions and
  parent {H}amiltonians},\ }\href {https://doi.org/10.1103/PhysRevB.94.245104}
  {\bibfield  {journal} {\bibinfo  {journal} {Phys. Rev. B}\ }\textbf {\bibinfo
  {volume} {94}},\ \bibinfo {pages} {245104} (\bibinfo {year}
  {2016})}\BibitemShut {NoStop}%
\bibitem [{\citenamefont {Liu}\ \emph {et~al.}(2018)\citenamefont {Liu},
  \citenamefont {Gromov},\ and\ \citenamefont {Papi\ifmmode~\acute{c}\else
  \'{c}\fi{}}}]{Liu_Gromov_Papic_PRB2018_FQHquench}%
  \BibitemOpen
  \bibfield  {author} {\bibinfo {author} {\bibfnamefont {Z.}~\bibnamefont
  {Liu}}, \bibinfo {author} {\bibfnamefont {A.}~\bibnamefont {Gromov}},\ and\
  \bibinfo {author} {\bibfnamefont {Z.}~\bibnamefont
  {Papi\ifmmode~\acute{c}\else \'{c}\fi{}}},\ }\bibfield  {title} {\bibinfo
  {title} {Geometric quench and nonequilibrium dynamics of fractional quantum
  {H}all states},\ }\href {https://doi.org/10.1103/PhysRevB.98.155140}
  {\bibfield  {journal} {\bibinfo  {journal} {Phys. Rev. B}\ }\textbf {\bibinfo
  {volume} {98}},\ \bibinfo {pages} {155140} (\bibinfo {year}
  {2018})}\BibitemShut {NoStop}%
\bibitem [{\citenamefont {Fremling}\ \emph {et~al.}(2018)\citenamefont
  {Fremling}, \citenamefont {Repellin}, \citenamefont {St{\'e}phan},
  \citenamefont {Moran}, \citenamefont {Slingerland},\ and\ \citenamefont
  {Haque}}]{Fremling_Haque_NJP2018_LLLdynamics}%
  \BibitemOpen
  \bibfield  {author} {\bibinfo {author} {\bibfnamefont {M.}~\bibnamefont
  {Fremling}}, \bibinfo {author} {\bibfnamefont {C.}~\bibnamefont {Repellin}},
  \bibinfo {author} {\bibfnamefont {J.-M.}\ \bibnamefont {St{\'e}phan}},
  \bibinfo {author} {\bibfnamefont {N.}~\bibnamefont {Moran}}, \bibinfo
  {author} {\bibfnamefont {J.}~\bibnamefont {Slingerland}},\ and\ \bibinfo
  {author} {\bibfnamefont {M.}~\bibnamefont {Haque}},\ }\bibfield  {title}
  {\bibinfo {title} {Dynamics and level statistics of interacting fermions in
  the lowest {L}andau level},\ }\href@noop {} {\bibfield  {journal} {\bibinfo
  {journal} {New Journal of Physics}\ }\textbf {\bibinfo {volume} {20}},\
  \bibinfo {pages} {103036} (\bibinfo {year} {2018})}\BibitemShut {NoStop}%
\bibitem [{\citenamefont {Diener}\ \emph {et~al.}(2003)\citenamefont {Diener},
  \citenamefont {Dudarev}, \citenamefont {Sundaram},\ and\ \citenamefont
  {Niu}}]{Niu_arXiV2003}%
  \BibitemOpen
  \bibfield  {author} {\bibinfo {author} {\bibfnamefont {R.~B.}\ \bibnamefont
  {Diener}}, \bibinfo {author} {\bibfnamefont {A.~M.}\ \bibnamefont {Dudarev}},
  \bibinfo {author} {\bibfnamefont {G.}~\bibnamefont {Sundaram}},\ and\
  \bibinfo {author} {\bibfnamefont {Q.}~\bibnamefont {Niu}},\ }\bibfield
  {title} {\bibinfo {title} {Intrinsic self-rotation of {BEC} {B}loch wave
  packets},\ }\href@noop {} {\bibfield  {journal} {\bibinfo  {journal} {arXiv
  preprint cond-mat/0306184}\ } (\bibinfo {year} {2003})}\BibitemShut {NoStop}%
\bibitem [{\citenamefont {Price}\ and\ \citenamefont
  {Cooper}(2012)}]{Price_Cooper_PRA2012}%
  \BibitemOpen
  \bibfield  {author} {\bibinfo {author} {\bibfnamefont {H.~M.}\ \bibnamefont
  {Price}}\ and\ \bibinfo {author} {\bibfnamefont {N.~R.}\ \bibnamefont
  {Cooper}},\ }\bibfield  {title} {\bibinfo {title} {Mapping the {B}erry
  curvature from semiclassical dynamics in optical lattices},\ }\href
  {https://doi.org/10.1103/PhysRevA.85.033620} {\bibfield  {journal} {\bibinfo
  {journal} {Phys. Rev. A}\ }\textbf {\bibinfo {volume} {85}},\ \bibinfo
  {pages} {033620} (\bibinfo {year} {2012})}\BibitemShut {NoStop}%
\bibitem [{\citenamefont {Roy}\ \emph {et~al.}(2015)\citenamefont {Roy},
  \citenamefont {Grushin}, \citenamefont {Moessner},\ and\ \citenamefont
  {Haque}}]{Roy_Grushin_Moessner_Haque_PRA2015}%
  \BibitemOpen
  \bibfield  {author} {\bibinfo {author} {\bibfnamefont {S.}~\bibnamefont
  {Roy}}, \bibinfo {author} {\bibfnamefont {A.~G.}\ \bibnamefont {Grushin}},
  \bibinfo {author} {\bibfnamefont {R.}~\bibnamefont {Moessner}},\ and\
  \bibinfo {author} {\bibfnamefont {M.}~\bibnamefont {Haque}},\ }\bibfield
  {title} {\bibinfo {title} {Wave-packet dynamics on {C}hern-band lattices in a
  trap},\ }\href {https://doi.org/10.1103/PhysRevA.92.063626} {\bibfield
  {journal} {\bibinfo  {journal} {Phys. Rev. A}\ }\textbf {\bibinfo {volume}
  {92}},\ \bibinfo {pages} {063626} (\bibinfo {year} {2015})}\BibitemShut
  {NoStop}%
\bibitem [{\citenamefont {Mugel}\ \emph {et~al.}(2017)\citenamefont {Mugel},
  \citenamefont {Dauphin}, \citenamefont {Massignan}, \citenamefont {Tarruell},
  \citenamefont {Lewenstein}, \citenamefont {Lobo},\ and\ \citenamefont
  {Celi}}]{Dauphin_Massignan_Lewenstein_Lobo_SciPost2017}%
  \BibitemOpen
  \bibfield  {author} {\bibinfo {author} {\bibfnamefont {S.}~\bibnamefont
  {Mugel}}, \bibinfo {author} {\bibfnamefont {A.}~\bibnamefont {Dauphin}},
  \bibinfo {author} {\bibfnamefont {P.}~\bibnamefont {Massignan}}, \bibinfo
  {author} {\bibfnamefont {L.}~\bibnamefont {Tarruell}}, \bibinfo {author}
  {\bibfnamefont {M.}~\bibnamefont {Lewenstein}}, \bibinfo {author}
  {\bibfnamefont {C.}~\bibnamefont {Lobo}},\ and\ \bibinfo {author}
  {\bibfnamefont {A.}~\bibnamefont {Celi}},\ }\bibfield  {title} {\bibinfo
  {title} {{Measuring {C}hern numbers in {H}ofstadter strips}},\ }\href
  {https://doi.org/10.21468/SciPostPhys.3.2.012} {\bibfield  {journal}
  {\bibinfo  {journal} {SciPost Phys.}\ }\textbf {\bibinfo {volume} {3}},\
  \bibinfo {pages} {012} (\bibinfo {year} {2017})}\BibitemShut {NoStop}%
\bibitem [{\citenamefont {Killi}\ and\ \citenamefont
  {Paramekanti}(2012)}]{Killi_Paramekanti_PRA2012}%
  \BibitemOpen
  \bibfield  {author} {\bibinfo {author} {\bibfnamefont {M.}~\bibnamefont
  {Killi}}\ and\ \bibinfo {author} {\bibfnamefont {A.}~\bibnamefont
  {Paramekanti}},\ }\bibfield  {title} {\bibinfo {title} {Use of quantum
  quenches to probe the equilibrium current patterns of ultracold atoms in an
  optical lattice},\ }\href {https://doi.org/10.1103/PhysRevA.85.061606}
  {\bibfield  {journal} {\bibinfo  {journal} {Phys. Rev. A}\ }\textbf {\bibinfo
  {volume} {85}},\ \bibinfo {pages} {061606(R)} (\bibinfo {year}
  {2012})}\BibitemShut {NoStop}%
\bibitem [{\citenamefont {Killi}\ \emph {et~al.}(2012)\citenamefont {Killi},
  \citenamefont {Trotzky},\ and\ \citenamefont
  {Paramekanti}}]{Killi_Trotzky_Paramekanti_PRA2012}%
  \BibitemOpen
  \bibfield  {author} {\bibinfo {author} {\bibfnamefont {M.}~\bibnamefont
  {Killi}}, \bibinfo {author} {\bibfnamefont {S.}~\bibnamefont {Trotzky}},\
  and\ \bibinfo {author} {\bibfnamefont {A.}~\bibnamefont {Paramekanti}},\
  }\bibfield  {title} {\bibinfo {title} {Anisotropic quantum quench in the
  presence of frustration or background gauge fields: A probe of bulk currents
  and topological chiral edge modes},\ }\href
  {https://doi.org/10.1103/PhysRevA.86.063632} {\bibfield  {journal} {\bibinfo
  {journal} {Phys. Rev. A}\ }\textbf {\bibinfo {volume} {86}},\ \bibinfo
  {pages} {063632} (\bibinfo {year} {2012})}\BibitemShut {NoStop}%
\bibitem [{\citenamefont {Goldman}\ \emph {et~al.}(2013)\citenamefont
  {Goldman}, \citenamefont {Dalibard}, \citenamefont {Dauphin}, \citenamefont
  {Gerbier}, \citenamefont {Lewenstein}, \citenamefont {Zoller},\ and\
  \citenamefont
  {Spielman}}]{Goldman_Dalibard_Lewenstein_Zoller_Spielman_PNAS2013}%
  \BibitemOpen
  \bibfield  {author} {\bibinfo {author} {\bibfnamefont {N.}~\bibnamefont
  {Goldman}}, \bibinfo {author} {\bibfnamefont {J.}~\bibnamefont {Dalibard}},
  \bibinfo {author} {\bibfnamefont {A.}~\bibnamefont {Dauphin}}, \bibinfo
  {author} {\bibfnamefont {F.}~\bibnamefont {Gerbier}}, \bibinfo {author}
  {\bibfnamefont {M.}~\bibnamefont {Lewenstein}}, \bibinfo {author}
  {\bibfnamefont {P.}~\bibnamefont {Zoller}},\ and\ \bibinfo {author}
  {\bibfnamefont {I.~B.}\ \bibnamefont {Spielman}},\ }\bibfield  {title}
  {\bibinfo {title} {Direct imaging of topological edge states in cold-atom
  systems},\ }\href {https://doi.org/10.1073/pnas.1300170110} {\bibfield
  {journal} {\bibinfo  {journal} {Proceedings of the National Academy of
  Sciences}\ }\textbf {\bibinfo {volume} {110}},\ \bibinfo {pages} {6736}
  (\bibinfo {year} {2013})}\BibitemShut {NoStop}%
\bibitem [{\citenamefont {Dauphin}\ and\ \citenamefont
  {Goldman}(2013)}]{Dauphin_Goldman_PRL2013}%
  \BibitemOpen
  \bibfield  {author} {\bibinfo {author} {\bibfnamefont {A.}~\bibnamefont
  {Dauphin}}\ and\ \bibinfo {author} {\bibfnamefont {N.}~\bibnamefont
  {Goldman}},\ }\bibfield  {title} {\bibinfo {title} {Extracting the {C}hern
  number from the dynamics of a {F}ermi gas: Implementing a quantum {H}all bar
  for cold atoms},\ }\href {https://doi.org/10.1103/PhysRevLett.111.135302}
  {\bibfield  {journal} {\bibinfo  {journal} {Phys. Rev. Lett.}\ }\textbf
  {\bibinfo {volume} {111}},\ \bibinfo {pages} {135302} (\bibinfo {year}
  {2013})}\BibitemShut {NoStop}%
\bibitem [{\citenamefont {Hauke}\ \emph {et~al.}(2014)\citenamefont {Hauke},
  \citenamefont {Lewenstein},\ and\ \citenamefont
  {Eckardt}}]{Hauke_Lewenstein_Eckardt_tomography_PRL2014}%
  \BibitemOpen
  \bibfield  {author} {\bibinfo {author} {\bibfnamefont {P.}~\bibnamefont
  {Hauke}}, \bibinfo {author} {\bibfnamefont {M.}~\bibnamefont {Lewenstein}},\
  and\ \bibinfo {author} {\bibfnamefont {A.}~\bibnamefont {Eckardt}},\
  }\bibfield  {title} {\bibinfo {title} {Tomography of band insulators from
  quench dynamics},\ }\href {https://doi.org/10.1103/PhysRevLett.113.045303}
  {\bibfield  {journal} {\bibinfo  {journal} {Phys. Rev. Lett.}\ }\textbf
  {\bibinfo {volume} {113}},\ \bibinfo {pages} {045303} (\bibinfo {year}
  {2014})}\BibitemShut {NoStop}%
\bibitem [{\citenamefont {Caio}\ \emph {et~al.}(2015)\citenamefont {Caio},
  \citenamefont {Cooper},\ and\ \citenamefont
  {Bhaseen}}]{Cooper_Bhaseen_quench_PRL2015}%
  \BibitemOpen
  \bibfield  {author} {\bibinfo {author} {\bibfnamefont {M.~D.}\ \bibnamefont
  {Caio}}, \bibinfo {author} {\bibfnamefont {N.~R.}\ \bibnamefont {Cooper}},\
  and\ \bibinfo {author} {\bibfnamefont {M.~J.}\ \bibnamefont {Bhaseen}},\
  }\bibfield  {title} {\bibinfo {title} {Quantum quenches in {C}hern
  insulators},\ }\href {https://doi.org/10.1103/PhysRevLett.115.236403}
  {\bibfield  {journal} {\bibinfo  {journal} {Phys. Rev. Lett.}\ }\textbf
  {\bibinfo {volume} {115}},\ \bibinfo {pages} {236403} (\bibinfo {year}
  {2015})}\BibitemShut {NoStop}%
\bibitem [{\citenamefont {Grushin}\ \emph {et~al.}(2016)\citenamefont
  {Grushin}, \citenamefont {Roy},\ and\ \citenamefont
  {Haque}}]{Grushin_Roy_Haque_JSM2015}%
  \BibitemOpen
  \bibfield  {author} {\bibinfo {author} {\bibfnamefont {A.~G.}\ \bibnamefont
  {Grushin}}, \bibinfo {author} {\bibfnamefont {S.}~\bibnamefont {Roy}},\ and\
  \bibinfo {author} {\bibfnamefont {M.}~\bibnamefont {Haque}},\ }\bibfield
  {title} {\bibinfo {title} {Response of fermions in {C}hern bands to spatially
  local quenches},\ }\href@noop {} {\bibfield  {journal} {\bibinfo  {journal}
  {Journal of Statistical Mechanics: Theory and Experiment}\ }\textbf {\bibinfo
  {volume} {2016}},\ \bibinfo {pages} {083103} (\bibinfo {year}
  {2016})}\BibitemShut {NoStop}%
\bibitem [{\citenamefont {Yu}(2017)}]{Yu_Haldanequench_phasevortices_PRA2017}%
  \BibitemOpen
  \bibfield  {author} {\bibinfo {author} {\bibfnamefont {J.}~\bibnamefont
  {Yu}},\ }\bibfield  {title} {\bibinfo {title} {Phase vortices of the quenched
  {H}aldane model},\ }\href {https://doi.org/10.1103/PhysRevA.96.023601}
  {\bibfield  {journal} {\bibinfo  {journal} {Phys. Rev. A}\ }\textbf {\bibinfo
  {volume} {96}},\ \bibinfo {pages} {023601} (\bibinfo {year}
  {2017})}\BibitemShut {NoStop}%
\bibitem [{\citenamefont {Mardanya}\ \emph {et~al.}(2018)\citenamefont
  {Mardanya}, \citenamefont {Bhattacharya}, \citenamefont {Agarwal},\ and\
  \citenamefont {Dutta}}]{Dutta_Haldanequench_edgecurrents_PRB2018}%
  \BibitemOpen
  \bibfield  {author} {\bibinfo {author} {\bibfnamefont {S.}~\bibnamefont
  {Mardanya}}, \bibinfo {author} {\bibfnamefont {U.}~\bibnamefont
  {Bhattacharya}}, \bibinfo {author} {\bibfnamefont {A.}~\bibnamefont
  {Agarwal}},\ and\ \bibinfo {author} {\bibfnamefont {A.}~\bibnamefont
  {Dutta}},\ }\bibfield  {title} {\bibinfo {title} {Dynamics of edge currents
  in a linearly quenched {H}aldane model},\ }\href
  {https://doi.org/10.1103/PhysRevB.97.115443} {\bibfield  {journal} {\bibinfo
  {journal} {Phys. Rev. B}\ }\textbf {\bibinfo {volume} {97}},\ \bibinfo
  {pages} {115443} (\bibinfo {year} {2018})}\BibitemShut {NoStop}%
\bibitem [{\citenamefont {Dong}\ \emph {et~al.}(2018)\citenamefont {Dong},
  \citenamefont {Grushin}, \citenamefont {Motruk},\ and\ \citenamefont
  {Pollmann}}]{Dong_Grushin_Motruk_Pollmann_PRL2018}%
  \BibitemOpen
  \bibfield  {author} {\bibinfo {author} {\bibfnamefont {X.-Y.}\ \bibnamefont
  {Dong}}, \bibinfo {author} {\bibfnamefont {A.~G.}\ \bibnamefont {Grushin}},
  \bibinfo {author} {\bibfnamefont {J.}~\bibnamefont {Motruk}},\ and\ \bibinfo
  {author} {\bibfnamefont {F.}~\bibnamefont {Pollmann}},\ }\bibfield  {title}
  {\bibinfo {title} {Charge excitation dynamics in bosonic fractional {C}hern
  insulators},\ }\href {https://doi.org/10.1103/PhysRevLett.121.086401}
  {\bibfield  {journal} {\bibinfo  {journal} {Phys. Rev. Lett.}\ }\textbf
  {\bibinfo {volume} {121}},\ \bibinfo {pages} {086401} (\bibinfo {year}
  {2018})}\BibitemShut {NoStop}%
\bibitem [{\citenamefont {Atala}\ \emph {et~al.}(2013)\citenamefont {Atala},
  \citenamefont {Aidelsburger}, \citenamefont {Barreiro}, \citenamefont
  {Abanin}, \citenamefont {Kitagawa}, \citenamefont {Demler},\ and\
  \citenamefont
  {Bloch}}]{Aidelsburger_Abanin_Demler_Bloch_Zakphase_NatPhys2013}%
  \BibitemOpen
  \bibfield  {author} {\bibinfo {author} {\bibfnamefont {M.}~\bibnamefont
  {Atala}}, \bibinfo {author} {\bibfnamefont {M.}~\bibnamefont {Aidelsburger}},
  \bibinfo {author} {\bibfnamefont {J.~T.}\ \bibnamefont {Barreiro}}, \bibinfo
  {author} {\bibfnamefont {D.}~\bibnamefont {Abanin}}, \bibinfo {author}
  {\bibfnamefont {T.}~\bibnamefont {Kitagawa}}, \bibinfo {author}
  {\bibfnamefont {E.}~\bibnamefont {Demler}},\ and\ \bibinfo {author}
  {\bibfnamefont {I.}~\bibnamefont {Bloch}},\ }\bibfield  {title} {\bibinfo
  {title} {Direct measurement of the {Z}ak phase in topological {B}loch
  bands},\ }\href@noop {} {\bibfield  {journal} {\bibinfo  {journal} {Nature
  Physics}\ }\textbf {\bibinfo {volume} {9}},\ \bibinfo {pages} {795} (\bibinfo
  {year} {2013})}\BibitemShut {NoStop}%
\bibitem [{\citenamefont {Jotzu}\ \emph {et~al.}(2014)\citenamefont {Jotzu},
  \citenamefont {Messer}, \citenamefont {Desbuquois}, \citenamefont {Lebrat},
  \citenamefont {Uehlinger}, \citenamefont {Greif},\ and\ \citenamefont
  {Esslinger}}]{Esslingergroup_Haldanemodel_Nature2014}%
  \BibitemOpen
  \bibfield  {author} {\bibinfo {author} {\bibfnamefont {G.}~\bibnamefont
  {Jotzu}}, \bibinfo {author} {\bibfnamefont {M.}~\bibnamefont {Messer}},
  \bibinfo {author} {\bibfnamefont {R.}~\bibnamefont {Desbuquois}}, \bibinfo
  {author} {\bibfnamefont {M.}~\bibnamefont {Lebrat}}, \bibinfo {author}
  {\bibfnamefont {T.}~\bibnamefont {Uehlinger}}, \bibinfo {author}
  {\bibfnamefont {D.}~\bibnamefont {Greif}},\ and\ \bibinfo {author}
  {\bibfnamefont {T.}~\bibnamefont {Esslinger}},\ }\bibfield  {title} {\bibinfo
  {title} {Experimental realization of the topological {H}aldane model with
  ultracold fermions},\ }\href@noop {} {\bibfield  {journal} {\bibinfo
  {journal} {Nature}\ }\textbf {\bibinfo {volume} {515}},\ \bibinfo {pages}
  {237} (\bibinfo {year} {2014})}\BibitemShut {NoStop}%
\bibitem [{\citenamefont {Aidelsburger}\ \emph {et~al.}(2015)\citenamefont
  {Aidelsburger}, \citenamefont {Lohse}, \citenamefont {Schweizer},
  \citenamefont {Atala}, \citenamefont {Barreiro}, \citenamefont
  {Nascimb{\`e}ne}, \citenamefont {Cooper}, \citenamefont {Bloch},\ and\
  \citenamefont {Goldman}}]{Aidelsberger_Cooper_Bloch_Goldman_NatPhys2015}%
  \BibitemOpen
  \bibfield  {author} {\bibinfo {author} {\bibfnamefont {M.}~\bibnamefont
  {Aidelsburger}}, \bibinfo {author} {\bibfnamefont {M.}~\bibnamefont {Lohse}},
  \bibinfo {author} {\bibfnamefont {C.}~\bibnamefont {Schweizer}}, \bibinfo
  {author} {\bibfnamefont {M.}~\bibnamefont {Atala}}, \bibinfo {author}
  {\bibfnamefont {J.~T.}\ \bibnamefont {Barreiro}}, \bibinfo {author}
  {\bibfnamefont {S.}~\bibnamefont {Nascimb{\`e}ne}}, \bibinfo {author}
  {\bibfnamefont {N.}~\bibnamefont {Cooper}}, \bibinfo {author} {\bibfnamefont
  {I.}~\bibnamefont {Bloch}},\ and\ \bibinfo {author} {\bibfnamefont
  {N.}~\bibnamefont {Goldman}},\ }\bibfield  {title} {\bibinfo {title}
  {Measuring the {C}hern number of {H}ofstadter bands with ultracold bosonic
  atoms},\ }\href {https://doi.org/10.1038/nphys3171} {\bibfield  {journal}
  {\bibinfo  {journal} {Nature Physics}\ }\textbf {\bibinfo {volume} {11}},\
  \bibinfo {pages} {162} (\bibinfo {year} {2015})}\BibitemShut {NoStop}%
\bibitem [{\citenamefont {Duca}\ \emph {et~al.}(2015)\citenamefont {Duca},
  \citenamefont {Li}, \citenamefont {Reitter}, \citenamefont {Bloch},
  \citenamefont {Schleier-Smith},\ and\ \citenamefont
  {Schneider}}]{Bloch_Schneider_interferometer_Science2015}%
  \BibitemOpen
  \bibfield  {author} {\bibinfo {author} {\bibfnamefont {L.}~\bibnamefont
  {Duca}}, \bibinfo {author} {\bibfnamefont {T.}~\bibnamefont {Li}}, \bibinfo
  {author} {\bibfnamefont {M.}~\bibnamefont {Reitter}}, \bibinfo {author}
  {\bibfnamefont {I.}~\bibnamefont {Bloch}}, \bibinfo {author} {\bibfnamefont
  {M.}~\bibnamefont {Schleier-Smith}},\ and\ \bibinfo {author} {\bibfnamefont
  {U.}~\bibnamefont {Schneider}},\ }\bibfield  {title} {\bibinfo {title} {An
  {A}haronov-{B}ohm interferometer for determining {B}loch band topology},\
  }\href@noop {} {\bibfield  {journal} {\bibinfo  {journal} {Science}\ }\textbf
  {\bibinfo {volume} {347}},\ \bibinfo {pages} {288} (\bibinfo {year}
  {2015})}\BibitemShut {NoStop}%
\bibitem [{\citenamefont {Stuhl}\ \emph {et~al.}(2015)\citenamefont {Stuhl},
  \citenamefont {Lu}, \citenamefont {Aycock}, \citenamefont {Genkina},\ and\
  \citenamefont {Spielman}}]{Spielman_visualizing_edgestates_Science2015}%
  \BibitemOpen
  \bibfield  {author} {\bibinfo {author} {\bibfnamefont {B.}~\bibnamefont
  {Stuhl}}, \bibinfo {author} {\bibfnamefont {H.-I.}\ \bibnamefont {Lu}},
  \bibinfo {author} {\bibfnamefont {L.}~\bibnamefont {Aycock}}, \bibinfo
  {author} {\bibfnamefont {D.}~\bibnamefont {Genkina}},\ and\ \bibinfo {author}
  {\bibfnamefont {I.}~\bibnamefont {Spielman}},\ }\bibfield  {title} {\bibinfo
  {title} {Visualizing edge states with an atomic {B}ose gas in the quantum
  {H}all regime},\ }\href@noop {} {\bibfield  {journal} {\bibinfo  {journal}
  {Science}\ }\textbf {\bibinfo {volume} {349}},\ \bibinfo {pages} {1514}
  (\bibinfo {year} {2015})}\BibitemShut {NoStop}%
\bibitem [{\citenamefont {Fl{\"a}schner}\ \emph {et~al.}(2016)\citenamefont
  {Fl{\"a}schner}, \citenamefont {Rem}, \citenamefont {Tarnowski},
  \citenamefont {Vogel}, \citenamefont {L{\"u}hmann}, \citenamefont
  {Sengstock},\ and\ \citenamefont
  {Weitenberg}}]{Sengstock_Weitenberg_Berrycurvature_Floquet_2016experimental}%
  \BibitemOpen
  \bibfield  {author} {\bibinfo {author} {\bibfnamefont {N.}~\bibnamefont
  {Fl{\"a}schner}}, \bibinfo {author} {\bibfnamefont {B.}~\bibnamefont {Rem}},
  \bibinfo {author} {\bibfnamefont {M.}~\bibnamefont {Tarnowski}}, \bibinfo
  {author} {\bibfnamefont {D.}~\bibnamefont {Vogel}}, \bibinfo {author}
  {\bibfnamefont {D.-S.}\ \bibnamefont {L{\"u}hmann}}, \bibinfo {author}
  {\bibfnamefont {K.}~\bibnamefont {Sengstock}},\ and\ \bibinfo {author}
  {\bibfnamefont {C.}~\bibnamefont {Weitenberg}},\ }\bibfield  {title}
  {\bibinfo {title} {Experimental reconstruction of the {B}erry curvature in a
  {F}loquet {B}loch band},\ }\href@noop {} {\bibfield  {journal} {\bibinfo
  {journal} {Science}\ }\textbf {\bibinfo {volume} {352}},\ \bibinfo {pages}
  {1091} (\bibinfo {year} {2016})}\BibitemShut {NoStop}%
\bibitem [{\citenamefont {Sun}\ \emph {et~al.}(2018)\citenamefont {Sun},
  \citenamefont {Yi}, \citenamefont {Wang}, \citenamefont {Zhang},
  \citenamefont {Sanders}, \citenamefont {Xu}, \citenamefont {Wang},
  \citenamefont {Schmiedmayer}, \citenamefont {Deng}, \citenamefont {Liu},
  \citenamefont {Chen},\ and\ \citenamefont
  {Pan}}]{Pan_Shanghai_experiment_PRL2018}%
  \BibitemOpen
  \bibfield  {author} {\bibinfo {author} {\bibfnamefont {W.}~\bibnamefont
  {Sun}}, \bibinfo {author} {\bibfnamefont {C.-R.}\ \bibnamefont {Yi}},
  \bibinfo {author} {\bibfnamefont {B.-Z.}\ \bibnamefont {Wang}}, \bibinfo
  {author} {\bibfnamefont {W.-W.}\ \bibnamefont {Zhang}}, \bibinfo {author}
  {\bibfnamefont {B.~C.}\ \bibnamefont {Sanders}}, \bibinfo {author}
  {\bibfnamefont {X.-T.}\ \bibnamefont {Xu}}, \bibinfo {author} {\bibfnamefont
  {Z.-Y.}\ \bibnamefont {Wang}}, \bibinfo {author} {\bibfnamefont
  {J.}~\bibnamefont {Schmiedmayer}}, \bibinfo {author} {\bibfnamefont
  {Y.}~\bibnamefont {Deng}}, \bibinfo {author} {\bibfnamefont {X.-J.}\
  \bibnamefont {Liu}}, \bibinfo {author} {\bibfnamefont {S.}~\bibnamefont
  {Chen}},\ and\ \bibinfo {author} {\bibfnamefont {J.-W.}\ \bibnamefont
  {Pan}},\ }\bibfield  {title} {\bibinfo {title} {Uncover topology by quantum
  quench dynamics},\ }\href {https://doi.org/10.1103/PhysRevLett.121.250403}
  {\bibfield  {journal} {\bibinfo  {journal} {Phys. Rev. Lett.}\ }\textbf
  {\bibinfo {volume} {121}},\ \bibinfo {pages} {250403} (\bibinfo {year}
  {2018})}\BibitemShut {NoStop}%
\bibitem [{\citenamefont {Tarnowski}\ \emph {et~al.}(2019)\citenamefont
  {Tarnowski}, \citenamefont {{\"U}nal}, \citenamefont {Fl{\"a}schner},
  \citenamefont {Rem}, \citenamefont {Eckardt}, \citenamefont {Sengstock},\
  and\ \citenamefont
  {Weitenberg}}]{Eckardt_Sengstock_Wittenberg_topologyfromdynamics_NatCom2019}%
  \BibitemOpen
  \bibfield  {author} {\bibinfo {author} {\bibfnamefont {M.}~\bibnamefont
  {Tarnowski}}, \bibinfo {author} {\bibfnamefont {F.~N.}\ \bibnamefont
  {{\"U}nal}}, \bibinfo {author} {\bibfnamefont {N.}~\bibnamefont
  {Fl{\"a}schner}}, \bibinfo {author} {\bibfnamefont {B.~S.}\ \bibnamefont
  {Rem}}, \bibinfo {author} {\bibfnamefont {A.}~\bibnamefont {Eckardt}},
  \bibinfo {author} {\bibfnamefont {K.}~\bibnamefont {Sengstock}},\ and\
  \bibinfo {author} {\bibfnamefont {C.}~\bibnamefont {Weitenberg}},\ }\bibfield
   {title} {\bibinfo {title} {Measuring topology from dynamics by obtaining the
  {C}hern number from a linking number},\ }\href@noop {} {\bibfield  {journal}
  {\bibinfo  {journal} {Nature Communications}\ }\textbf {\bibinfo {volume}
  {10}},\ \bibinfo {pages} {1728} (\bibinfo {year} {2019})}\BibitemShut
  {NoStop}%
\bibitem [{\citenamefont {Nandy}\ \emph {et~al.}(2019)\citenamefont {Nandy},
  \citenamefont {Srivatsa},\ and\ \citenamefont
  {Nielsen}}]{Nandy_Nielsen_truncation_PRB2019}%
  \BibitemOpen
  \bibfield  {author} {\bibinfo {author} {\bibfnamefont {D.~K.}\ \bibnamefont
  {Nandy}}, \bibinfo {author} {\bibfnamefont {N.~S.}\ \bibnamefont
  {Srivatsa}},\ and\ \bibinfo {author} {\bibfnamefont {A.~E.~B.}\ \bibnamefont
  {Nielsen}},\ }\bibfield  {title} {\bibinfo {title} {Truncation of lattice
  fractional quantum {H}all {H}amiltonians derived from conformal field
  theory},\ }\href {https://doi.org/10.1103/PhysRevB.100.035123} {\bibfield
  {journal} {\bibinfo  {journal} {Phys. Rev. B}\ }\textbf {\bibinfo {volume}
  {100}},\ \bibinfo {pages} {035123} (\bibinfo {year} {2019})}\BibitemShut
  {NoStop}%
\bibitem [{\citenamefont {Chesnokov}\ and\ \citenamefont
  {Kolovsky}(2014)}]{Kolovsky_EPL2014}%
  \BibitemOpen
  \bibfield  {author} {\bibinfo {author} {\bibfnamefont {I.~Y.}\ \bibnamefont
  {Chesnokov}}\ and\ \bibinfo {author} {\bibfnamefont {A.~R.}\ \bibnamefont
  {Kolovsky}},\ }\bibfield  {title} {\bibinfo {title} {{L}andau-stark states in
  finite lattices and edge-induced {B}loch oscillations},\ }\href@noop {}
  {\bibfield  {journal} {\bibinfo  {journal} {EPL (Europhysics Letters)}\
  }\textbf {\bibinfo {volume} {106}},\ \bibinfo {pages} {50001} (\bibinfo
  {year} {2014})}\BibitemShut {NoStop}%
\bibitem [{\citenamefont {Kolovsky}\ \emph {et~al.}(2014)\citenamefont
  {Kolovsky}, \citenamefont {Grusdt},\ and\ \citenamefont
  {Fleischhauer}}]{Kolovsky_Grusdt_Fleischhauer_PRA2014}%
  \BibitemOpen
  \bibfield  {author} {\bibinfo {author} {\bibfnamefont {A.~R.}\ \bibnamefont
  {Kolovsky}}, \bibinfo {author} {\bibfnamefont {F.}~\bibnamefont {Grusdt}},\
  and\ \bibinfo {author} {\bibfnamefont {M.}~\bibnamefont {Fleischhauer}},\
  }\bibfield  {title} {\bibinfo {title} {Quantum particle in a parabolic
  lattice in the presence of a gauge field},\ }\href
  {https://doi.org/10.1103/PhysRevA.89.033607} {\bibfield  {journal} {\bibinfo
  {journal} {Phys. Rev. A}\ }\textbf {\bibinfo {volume} {89}},\ \bibinfo
  {pages} {033607} (\bibinfo {year} {2014})}\BibitemShut {NoStop}%
\bibitem [{\citenamefont {Haldane}(1988)}]{Haldane_PRL1988}%
  \BibitemOpen
  \bibfield  {author} {\bibinfo {author} {\bibfnamefont {F.~D.~M.}\
  \bibnamefont {Haldane}},\ }\bibfield  {title} {\bibinfo {title} {Model for a
  quantum {H}all effect without {L}andau levels: Condensed-matter realization
  of the "parity anomaly"},\ }\href
  {https://doi.org/10.1103/PhysRevLett.61.2015} {\bibfield  {journal} {\bibinfo
   {journal} {Phys. Rev. Lett.}\ }\textbf {\bibinfo {volume} {61}},\ \bibinfo
  {pages} {2015} (\bibinfo {year} {1988})}\BibitemShut {NoStop}%
\bibitem [{\citenamefont {Rechtsman}\ \emph {et~al.}(2013)\citenamefont
  {Rechtsman}, \citenamefont {Zeuner}, \citenamefont {Plotnik}, \citenamefont
  {Lumer}, \citenamefont {Podolsky}, \citenamefont {Dreisow}, \citenamefont
  {Nolte}, \citenamefont {Segev},\ and\ \citenamefont
  {Szameit}}]{Szameit_photonic_topological_expt_Nature2013}%
  \BibitemOpen
  \bibfield  {author} {\bibinfo {author} {\bibfnamefont {M.~C.}\ \bibnamefont
  {Rechtsman}}, \bibinfo {author} {\bibfnamefont {J.~M.}\ \bibnamefont
  {Zeuner}}, \bibinfo {author} {\bibfnamefont {Y.}~\bibnamefont {Plotnik}},
  \bibinfo {author} {\bibfnamefont {Y.}~\bibnamefont {Lumer}}, \bibinfo
  {author} {\bibfnamefont {D.}~\bibnamefont {Podolsky}}, \bibinfo {author}
  {\bibfnamefont {F.}~\bibnamefont {Dreisow}}, \bibinfo {author} {\bibfnamefont
  {S.}~\bibnamefont {Nolte}}, \bibinfo {author} {\bibfnamefont
  {M.}~\bibnamefont {Segev}},\ and\ \bibinfo {author} {\bibfnamefont
  {A.}~\bibnamefont {Szameit}},\ }\bibfield  {title} {\bibinfo {title}
  {Photonic {F}loquet topological insulators},\ }\href@noop {} {\bibfield
  {journal} {\bibinfo  {journal} {Nature}\ }\textbf {\bibinfo {volume} {496}},\
  \bibinfo {pages} {196} (\bibinfo {year} {2013})}\BibitemShut {NoStop}%
\end{thebibliography}
\end{document}